\documentclass[twocolumn]{aastex631}

\usepackage[T1]{fontenc}
\usepackage{graphicx}	% Including figure files
\usepackage{amsmath}	% Advanced maths commands
\usepackage{subfigure}
\usepackage{booktabs}
\usepackage{xcolor} % for comment color

\usepackage{xspace}
\newcommand{\satgen}{{\tt\string SatGen}\xspace}

\begin{document}

\title{Why artificial disruption is not a concern for current cosmological simulations}

\author[0000-0003-0303-4188]{Feihong He}
\affiliation{Department of Astronomy, School of Physics and Astronomy, Shanghai Jiao Tong University, Shanghai, 200240, China}
\affiliation{Key Laboratory for Particle Astrophysics and Cosmology (MOE), Shanghai 200240, China}
\affiliation{Shanghai Key Laboratory for Particle Physics and Cosmology, Shanghai 200240, China}

\author[0000-0002-8010-6715]{Jiaxin Han}
\affiliation{Department of Astronomy, School of Physics and Astronomy, Shanghai Jiao Tong University, Shanghai, 200240, China}
\affiliation{Key Laboratory for Particle Astrophysics and Cosmology (MOE), Shanghai 200240, China}
\affiliation{Shanghai Key Laboratory for Particle Physics and Cosmology, Shanghai 200240, China}

%\collaboration{20}{(AAS Journals Data Editors)}

\author[0000-0001-7890-4964]{Zhaozhou Li}
\affiliation{Centre for Astrophysics and Planetary Science, Racah Institute of Physics, The Hebrew University, Jerusalem, 91904, Israel}

\correspondingauthor{Jiaxin Han}
\email{fhtouma@sjtu.edu.cn(FH), jiaxin.han@sjtu.edu.cn(JH)}

%\altaffiliation{AASTeX v6+ programmer}

%\affiliation{AAS Director of Publishing}
%\affiliation{American Astronomical Society \\
%1667 K Street NW, Suite 800 \\



\begin{abstract}
Recent studies suggest that cold dark matter subhalos are hard to disrupt and almost all cases of subhalo disruption observed in numerical simulations are due to numerical effects. However, these findings primarily relied on idealized numerical experiments, which do not fully capture the realistic conditions of subhalo evolution within a hierarchical cosmological context. Based on the Aquarius simulations, we identify clear segregation in the population of surviving and disrupted subhalos, which corresponds to \emph{two distinct acquisition channels} of subhalos. We find that all of the first-order subhalos accreted after redshift 2 survive to the present time without suffering from artificial disruption. On the other hand, most of the disrupted subhalos are sub-subhalos accreted at high redshift. Unlike the first-order subhalos, sub-subhalos experience pre-processing and many of them are accreted through major mergers at high redshift, resulting in very high mass loss rates. We confirm these high mass loss rates are physical through both numerical experiments and semi-analytical modeling, thus supporting a physical origin for their rapid disappearance in the simulation. Even though we cannot verify whether these subhalos have fully disrupted or not, their extreme mass loss rates dictate that they can at most contribute a negligible fraction to the very low mass end of the subhalo mass function. We thus conclude that current state-of-the-art cosmological simulations have reliably resolved the subhalo population.
\end{abstract}


\defcitealias{Han16}{Han16}
\defcitealias{vdB18b}{BO18}
\defcitealias{Jiang21}{Jiang21}
\defcitealias{Green21}{Green21}
\defcitealias{Errani21}{Errani21}

\section{Introduction} \label{sec:intro}

%introduction to subhalo
In the hierarchical structure formation paradigm, small dark matter halos form earlier. Then they accrete the surrounding material and merge other small halos to build up a larger halo, while the accreted small halo survives as subhalos whose masses are gradually stripped by the tidal field of the host halo. A halo carrying a subhalo continues to merge, transforming the subhalo into a sub-subhalo~\citep{Ghigna98, Gao04b, Gao04a, Aquarius, Ludlow09, Giocoli10, Gao12}.  Through this process, layer by layer, they combine to ultimately form a hierarchical universe.

%the population of subhalo is sensitive to the nature of dark matter, the evolution of satellite galaxies, interaction with the stellar stream, boost of dark matter annihilation

The precise prediction of the distribution and properties of the subhalo can make them useful probes in many astrophysical subjects. For example, the properties of subhalo are sensitive to the nature of dark matter. If the dark matter is ``warm" rather than ``cold", the abundance of the low mass subhalo will be significantly reduced~\citep[e.g.,][]{Wang07, Lovell14, Bose17b, Stucker22, He23}. Subhalos composed of self-interacting dark matter(SIDM) will form a core-like density profile at its inner region, which is in contrast with the cuspy density profile predicted by the cold dark matter (CDM) subhalo model~\citep[e.g.,][]{Dav01, Rocha13, Vogelsberger14, Elbert15, Tulin18}. In observation, gravitational lensing can provide constraints on the distribution of subhalos, thereby potentially distinguishing the properties of dark matter particles~\citep[e.g.,][]{Li16, Inoue16, Nierenberg17, Dai18, Wang23}. Some models predict that the interaction between subhalo and the stellar stream will modify the morphology of the stream, generating structures like ``gap" and ``spur"~\citep[e.g.,][]{Carlberg12, Carlberg13, Carlberg18, Bonaca19}. The dark matter annihilation signal detection also depends on how the subhalo is distributed~\citep[e.g.,][]{Tasitsimoi02, Springel08b, Han12a, Huang12, Stref19}. Lastly, the subhalos or satellite galaxies can serve as dynamical tracers for their host halo mass (e.g., 
\citealt{2009MNRAS.399..812W, 2013MNRAS.429.3079M, 2014MNRAS.441.1513O, 2019ApJ...886...69L, 2020ApJ...894...10L,2020IAUS..353..109H,
2021MNRAS.505.3907L,2022MNRAS.514.5890L}). %\zzc{too many refs here?}

Based on cosmological simulations, previous studies have established a good understanding of the mass and spatial distribution of subhalos. Numerous models have been developed to mimic the evolution of subhalos, often starting with a Monte-Carlo realization of the extended Press-Schechter theory~\citep[][]{EPS} to generate a merger tree, known as the semi-analytical models (SAM). Once the initial orbit and structure are assigned to a subhalo at infall time in the merger tree, a SAM follows the non-linear physical evolution of the subhalo in the host halo potential, accounting for various processes including dynamical friction, tidal stripping and tidal heating~\citep{TB01, TB04, TB05a, Tb05b, Zentner05}, eventually leading to a prediction of the final distribution of the subhalos~\citep[e.g.,][]{Penarrubia05, Galacticus, Pullen14, Jiang16}. Starting from the statistical properties of the subhalo population, \citet{Han16}(hereafter \citetalias{Han16}) proposed an analytical model for the mass and spatial distribution of subhalos by linking their infall states and final states through mass evolution. These models, without exception, require knowledge of the mass loss process of a subhalo caused by tidal stripping. However, in numerical simulations, the fate of a subhalo's evolution in the tidal field has always been a controversial issue.

Since the concept of subhalo was first proposed, the issue of ``overmerging" in numerical simulations has been a concern. It was believed that due to insufficient resolution, the number of subhalos in simulations was less than the number of observed satellites \citep{Kampen95, Moore96, Moore98, Klypin99}. However, as the resolution increased, the number of subhalos also increased, and soon this issue was overshadowed by the 'missing satellite' problem~\citep{Klypin99b, Moore99}. Nowadays, thanks to improvements in computational power, state-of-the-art numerical simulations suggest that about half of the accreted subhalos are destroyed by tidal forces \citep{Han16, Jiang16}. Analytically, it has also been argued that a subhalo will fully disrupt once it has been stripped to a critical radius below which the total energy becomes unbound~\citet{Hayashi03}. This criterion has been widely used for handling subhalo disruption in SAM~\citep[e.g.,][]{Galacticus, Jiang16, Green21}.

However, some recent studies have questioned this widely accepted criterion. %\citet{vdB17} analysed the segregation of subhalo population in the \textit{Bolshoi} simulation~\citep{Bolshoi} and suggested that the disrupted subhalo in simulation can be divided into two channels: withering and artificial disruption. If all subhalo disruption events are attributed to numerical effects, the abundance of the subhalo mass function could be boosted \jx{has anyone claimed this before?:} by a factor of 2.
Based on idealized simulations, many researchers claim that subhalos are very resistant to disruption by tidal effects~\citep[e.g.,][]{Penarrubia08, vdB18a, Errani20, Errani21, Darkos20}. \citet{vdB18a} conducted a series of analytical calculations and idealized simulations, finding that even when the injected tidal heating energy far exceeds the binding energy of the subhalo, the remnant will still be bound after re-virialization. This result arises because the majority of tidal energy is absorbed by particles already in a low binding energy state. Consequently, they estimated that tidal stripping alone might not be potent enough to fully disrupt the subhalo, suggesting that numerical effects could be responsible for observed disruptions. Subsequently, \citet{vdB18b}(hereafter \citetalias{vdB18b}) conducted convergence tests in idealized numerical simulations and found that when the mass resolution and force resolution of the subhalo is sufficient, the subhalo will not disrupt. When the number of particles is relatively small, the $N$-body system will be destroyed prematurely due to the influence of discreteness noise. When the force softening length is insufficient, the subhalo will also disintegrate prematurely. 

Also using the idealized simulation, \citet{Errani20} found that insufficient spatial resolution will artificially turn a cuspy density profile to a cored one, leading to numerical subhalo disruption. Subsequently, \citet{Errani21} found that a subhalo with a cuspy profile is hard to disrupt and its evolutionary track has an asymptotic endpoint which depends on the host potential around its orbit. \citet{Green21}(hereafter \citetalias{Green21}) used the SAM model {\tt\string SatGen} published by \citet{Jiang21} to generate the subhalo mass and spatial distribution and compared their results with the \textit{Bolshoi} simulation. If no artificial disruption model is added in the stripping recipe of {\tt\string SatGen}, their fiducial subhalo mass function from {\tt\string SatGen} will be higher than that of the simulation by about $20\%$, which is far less than expected but still significant. 

In this work, we will investigate the origin of the subhalo disruption in a state-of-the-art cosmological simulation. Our results show that the surviving and apparently disrupted (referred to as `orphan' hereafter) subhalo populations in simulation tend to have different properties. Previous idealized experiments were based on minor merger events and static analytical potential. We find that subhalos that meet these conditions in numerical simulations have indeed survived to the present time. However, there are still a population of subhalos that do not meet these conditions and they consist of the majority of the disrupted subhalos in the simulation. In particular, we find that previous work has overlooked the role of sub-subhalo (high-order subhalo) in this process. Using idealized simulations as well as a SAM, we critically evaluate the fate of these objects and their influences on the subhalo distribution. Our results suggest that artificial disruption is not a main concern for the subhalo population with a particle number larger than 300. The incompleteness only occurs at the low-mass (about dozens of particle numbers) end of the subhalo mass function, which is consistent with the previous work on the convergence test of the subhalo mass function~\citep[e.g.,][]{Diemand04, Aquarius}.
%\zzc{here can also briefly mention our main conclusion that artificial disruption is not a main concern for cosmological simulations .}

This paper is structured as follows. In section.~\ref{sec:simulationsubhalo}, we investigate the origin of the disrupted subhalos in a high resolution cosmological simulation. In section~\ref{sec:idealsim}, we verify the mass evolution of the disrupted subhalos using idealized simulations. In section~\ref{sec:satgen}, we further investigate the fate of the disrupted subhalos using a semi-analytical model. We give some discussions in section~\ref{sec:discussion} and make conclusions in section~\ref{sec:conclusion}. %\jx{check this:} Throughout this paper, we define the virial radius as the radius enclosing an average density of 200 times the critical density of the universe.

\section{The origin of disrupted Subhalos in cosmological simulations}\label{sec:simulationsubhalo}

\subsection{The Aquarius Simulations}
In this work, we use the Aquarius simulations \citep{Aquarius} to study the evolution of subhalos within a cosmological context. The Aquarius Project comprises a set of zoom-in simulations focusing on Milky-Way-size halos, encompassing six halos in total (designated A to F). Each halo is simulated at a series of resolutions, with Level 1 being the highest resolution and Level 5 the lowest. Our analysis primarily focuses on halos run at Level 2, corresponding to a mass resolution of approximately $10^{4}h^{-1}M_{\odot}$.

We identify subhalos and construct the merger tree using \textsc{hbt+} \citep{HBT+}. By default, the mass of a subhalo, denoted as $m$, is defined as the total mass of the self-bound dark matter particles. %In some cases, we will use the $m_{200\rm{c}}$ to define the mass of the subhalo. 

To study the evolution of subhalos, it is essential to determine their initial status at the time of infall. One common approach is to define the initial status at the time when the subhalo reaches its peak mass during its evolution. This peak mass, denoted as $m_{\rm{peak}}$,  marks the subhalo's maximum mass before it starts to lose mass, and the corresponding time is denoted as $z_{\rm{peak}}$. Alternatively, one may define the initial condition at the accretion time when an isolated halo first becomes a subhalo due to a merger. This time is denoted as $z_{\rm{acc}}$, with the corresponding infall mass $m_{\rm{acc}}$. Typically, $z_{\rm{peak}}$ is higher than $z_{\rm{acc}}$ and $m_{\rm{peak}}$ is larger than $m_{\rm{acc}}$~\citep{Sifon23}. It is worth noting that a (sub)halo can still lose mass even while it is isolated.
%\zzc{in what case? is it because it is approaching a more massive neighbor?}

To ensure subhalos are well-resolved at infall time, we select a subhalo sample according to the peak mass, with  $10^{8}h^{-1}M_{\odot}<m_{\rm{peak}} < 10^{9}h^{-1}M_{\odot}$. This mass limit corresponds to a particle number of about $10^{4} - 10^{5}$. All selected subhalos are located within the virial radius $R_{200}$ of its host halo at $z=0$. The virial radius is defined as the radius within which the average density of the enclosed matter is 200 times the critical density of the universe. 

\subsection{Definition of Subhalo Disruption}

\begin{figure*}[ht!]
    \centering
    \includegraphics[scale=0.5]{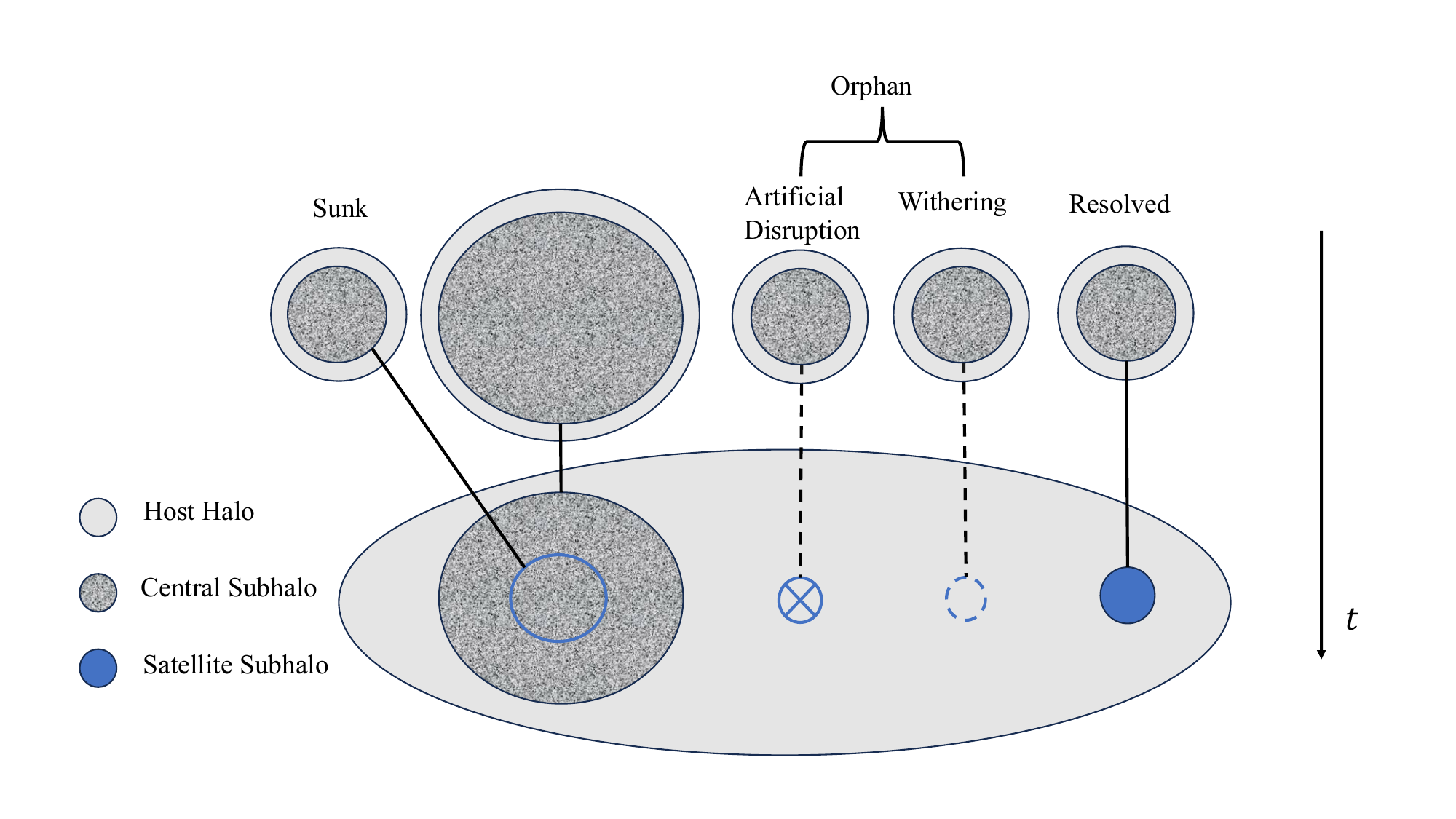}
    \caption{Classification of subhalos in a merger tree. The cosmic time flows from top to bottom. At the final snapshot of interest, a resolved subhalo has mass above the mass resolution (20 particles in our case), while orphans are those with no resolved descendant. The orphan class is further divided into artificial disruption (AD) and withering classes, according to whether the subhalo evolution truncates well above or close to the resolution limit. Sunk (or trapped) subhalos are those drifted to the center of another subhalo and stayed there therein without being disrupted first.
    }
    \label{fig:ill}
\end{figure*}

% The definition of subhalo disruption

%We want to discuss the nature of subhalo disruption. However, the essence of the disruption of the subhalo has not yet been truly determined. Our subhalo finder, \textsc{hbt+} will continue to trace the evolution of the subhalo. Even if one subhalo loses all of its mass, it can be labelled by the most bound particle before its mass under the mass resolution limit. For simplicity, we defined the subhalos with a mass loss down to the mass resolution limit $N_{\rm{min}}=20$ as the disrupted population. And subhalos still carry more than $N_{\rm{min}}$ particles at $z=0$ as the survival population. 

%The essence and definition of subhalo disruption were not clearly articulated.
In our subhalo finder, \textsc{hbt+}, the subhalo population can be divided into three distinct populations. The first population includes subhalos with a self-bound particle number greater than the given limit $N_{\rm{min}}=20$ at the time of interest. We will call these objects resolved subhalos. When the number of bound particles in a subhalo falls below $N_{\rm{min}}=20$, \textsc{hbt+} continues to track its most-bound particle till the end of the simulation. We classify these subhalos resolved with only one particle as the orphan population. In the following, we focus on classifying the subhalos according to their bound mass at $z=0$. Besides these two populations, \textsc{hbt+} identifies a third population of subhalos known as trapped (or sunk) subhalos. These occur when a subhalo has sunk to the center of another subhalo and becomes indistinguishable from the other in phase space. At the resolution level of the simulation, the two subhalos are identified as one merged subhalo. %We will briefly mention these in the subsection but will not delve deeply into their discussion. For more detailed discussions, please refer to \citet{HBT+}.

Following \citet{vdB17}, the orphan population can be further divided into two classes: \textit{withering} and \textit{disruption}. We define withering as the population that terminates their mass evolution around the mass resolution (20 particles) of the subhalo catalog so that it is likely they can continue to be resolved at a better resolution. On the other hand, disruption refers to subhalos that abruptly become unresolved at a final mass well above the 20-particle limit.  \citet{vdB18a} points out that tidal heating is insufficient to completely destroy the structure of CDM subhalos. Similarly, other studies, such as those by \citet{Amorisco21}, suggest that subhalos with cuspy density profiles can survive even when stripped to very low masses, implying that physical disruption is not prevalent. %Nevertheless, we still keep the classification of disruption to refer to a potential population of physically disintegrated subhalos that stay disrupted at infinite numerical resolution, even though their contribution may be null in the CDM universe. Note that any artificial disruption due to insufficient resolution will be classified as withering in our definition.  
Thus, the second channel, \textit{disruption}, actually refers to \textit{Artificial Disruption}(AD). Due to the finite time resolution of the mass evolution histories, it is possible that some subhalos classified as disruption are withering ones with a very high mass loss rate, so that they drop to the mass resolution before reaching the next snapshot. It is also possible that some withering subhalos actually stay unresolved when the mass resolution increases, i.e., become disrupted. Thus there is always some ambiguity in completely distinguishing the two classes. 

These classifications are summarized in Figure~\ref{fig:ill}. As it is difficult to completely distinguish between withering and disruption in our sample, in the following we focus on studying these two populations together under the orphan class, along with resolved and sunk classes. We discuss the distinction between withering and disruption in section~\ref{sec:ad} and section~\ref{sec:satgen}

\subsection{Dissecting the Subhalo Population}

% disrupted subhalo and survival subhalo population could be segregated

%To discuss the non-linear evolution of the subhalo, we should clarify which properties of the subhalo evolve with time. The most common one is the subhalo mass evolution. %It has been studied for many years. But overall, these studies are at a relatively macro level, such as the evolution of subhalo mass functions. Specifically, there are not many studies on the one-to-one correspondence of each subhalo.

Previous studies have attempted to separate resolved and orphan subhalos according to their initial masses~\citep[e.g.,][]{vdB17}.
%Previous studies on the distribution and evolutionary properties of subhalos have mostly focused on the initial subhalo mass. 
However, the unevolved subhalo mass functions for both populations take on the same shape, making it impossible to distinguish the two using their infall masses~\citep{Han12b, Jiang21}. In the following we explore the segregation of the two populations using some other properties.

\begin{figure}
    \centering
    \includegraphics[width=\columnwidth]{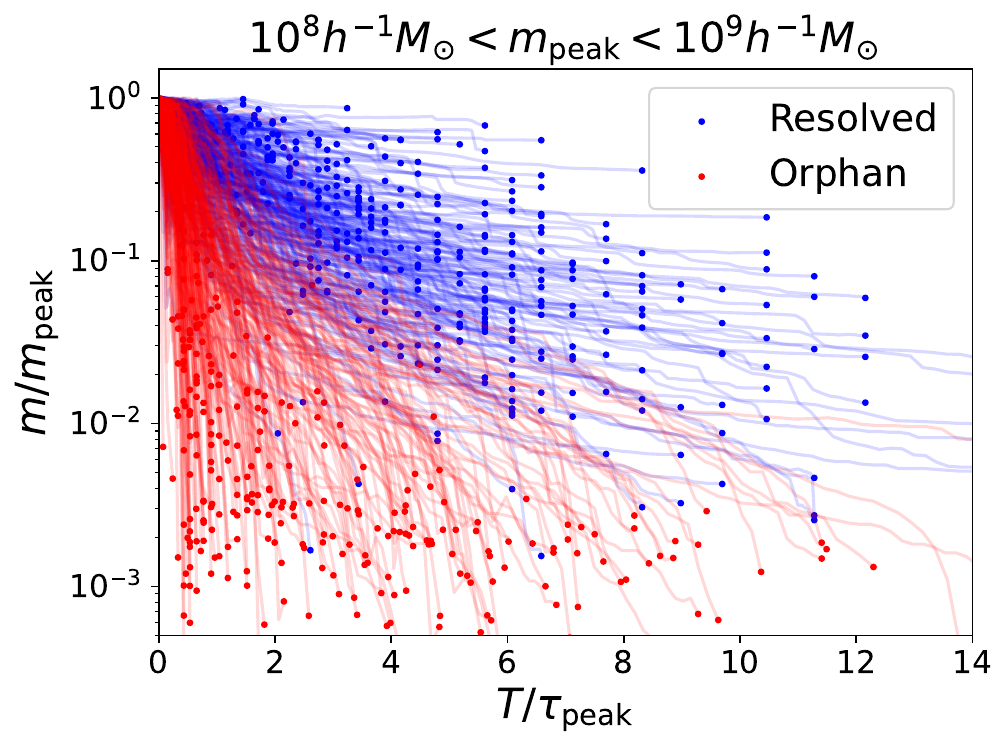}
    \caption{The mass evolution of subhalos in our sample. The x-axis is the evolution time starting from the peak-mass time of each subhalo, normalized by the dynamical time scale $\tau=2 \pi R_{\rm{h}}/V_{\rm{h}}$ of the host halo at the peak-mass time. The blue and red lines represent the resolved and orphan populations respectively. The solid dots mark the final state of each subhalo at its last resolvable snapshot.
    }
    \label{fig:scaled_massloss}
\end{figure}

Figure~\ref{fig:scaled_massloss} shows the normalized mass evolution histories of the subhalos in our sample. 
%Considering the dynamical time scale will change across cosmic time, a better way to study the mass loss curve is to scale the evolution time by a dynamical time scale $\tau$, which is defined as $\tau = 2 \pi R_{\rm{h}}/V_{\rm{h}}$. Here $R_{\rm{h}}$ is the radius of the host halo, and $V_{\rm{h}}=\sqrt{GM_{\rm{h}}/R_{\rm{h}}}$, $M_{\rm{h}}$ is the host halo mass. Here, we present the scaled mass loss curve, as shown in Fig.~\ref{fig:scaled_massloss}. 
It is evident that the orphan and surviving subhalo populations form two distinct groups. Surviving subhalos continue to lose mass until the present day, while the orphan population loses most of their masses in a short period. %The orphan population continues to lose material over several orbital periods, indicating that this is not merely a result of the shorter timescales in the early universe. 

\subsubsection{Segregating subhalos in their properties}\label{sec:property_segregation}

\begin{figure*}
    \centering
    \includegraphics[scale=0.5]{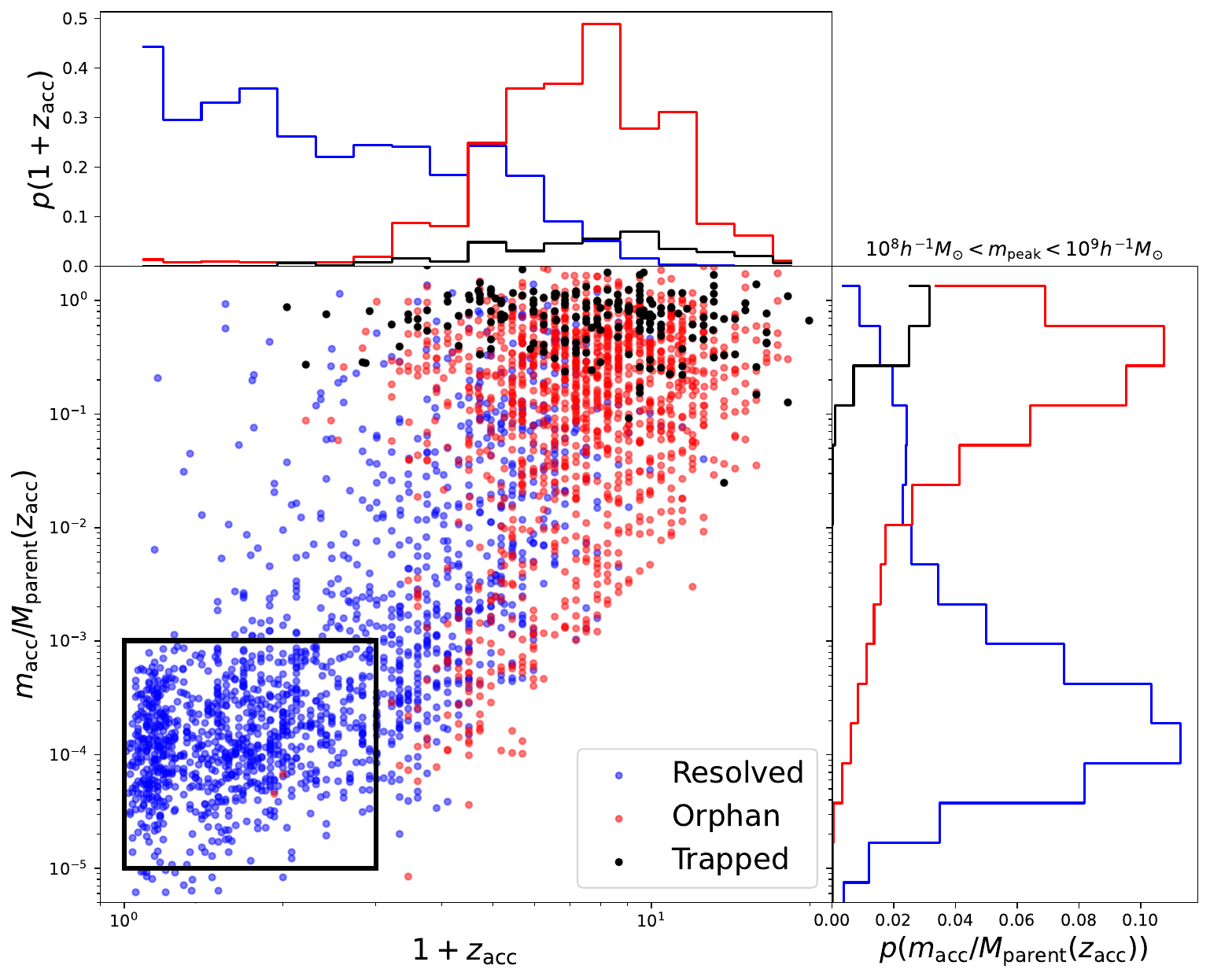}
    \caption{The joint distribution of the subhalo accretion time and its merger ratio at the accretion time, colored according to the subhalo classes as labeled. The side panels show the marginalized probability distributions of each population. The black rectangle shows the conditions studied in \citetalias{vdB18b}.
    %\zzc{use wider binning for the subpanels (it is too jumpy now). or kde for a smooth distribution.} \fhc{down.}
    }
    \label{fig:massratio_Zacc}
\end{figure*}

To pin down the physical origins for the different evolution paths of subhalos, in Figure~\ref{fig:massratio_Zacc} we show the joint distribution of subhalos in their accretion time, $z_{\rm acc}$, and merger mass ratio at $z_{\rm acc}$. Note the parent halo mass, $M_{\rm parent}$, refers to the mass of the halo that the subhalo merge or will merge with at $z_{\rm acc}$, which differs from the final host halo for high order subhalos.

The three subhalo populations are clearly segregated in Figure~\ref{fig:massratio_Zacc}. In terms of accretion time, orphan and trapped subhalos are mostly accreted at high redshift, while the resolved population are mostly accreted at low redshift. Almost all subhalos accreted after $z=2$ survive to the present day. In terms of merger ratio, the resolved population are mostly accreted through minor mergers with a mass ratio below $10^{-3}$. By contrast, the orphan population tend to have higher merger mass ratios, while almost all of the trapped populations are accreted through major mergers with very high mass ratios. %We can notice an envelope curve in Fig.~\ref{fig:massratio_Zacc}. It is the lower limit of the merger ratio for each redshift, which is $m_{\rm{min}}/M_{\rm{host, max}}(z)$, with $m_{\rm{min}}$ being the lowest subhalo mass in our sample and $M_{\rm{host, max}}(z)$ is the maximum host halo mass at each redshift.

The segregation of subhalos in Figure~\ref{fig:massratio_Zacc} is closely related to the hierarchical assembly mode of subhalos. To see this, in Table~\ref{tab:order_count} we show the fractional contributions to the subhalo count from different channels for subhalos of different orders. It can be clearly seen that the fate of a subhalo depends sensitively on its order. About $83\%$ of first-order subhalos are resolved while only $17\%$ first-order subhalos belong to the orphan population. For high-order subhalos, the resolved rate decreases dramatically compared with first-order subhalos. With the order increasing from 2 to 4, the resolved rate decreases from about $28\%$ to $16\%$. For comparison, we find a resolved fraction of about $53\%$ for the full subhalo sample, consistent with the results of \citetalias{Han16}.

\begin{figure*}
    \centering
    \begin{subfigure}
        \centering
        \includegraphics[scale=0.28]{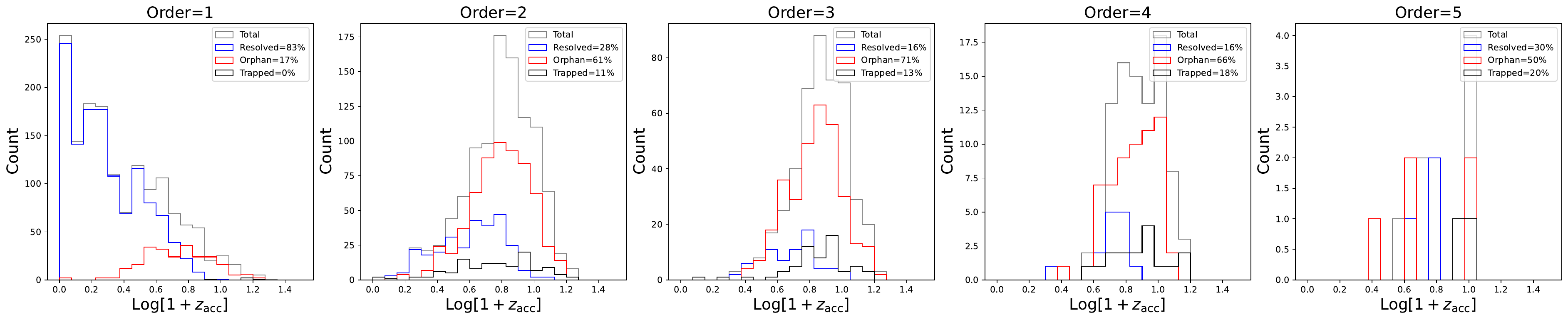}
    \end{subfigure}
    \centering
    \begin{subfigure}
        \centering
        \includegraphics[scale=0.28]{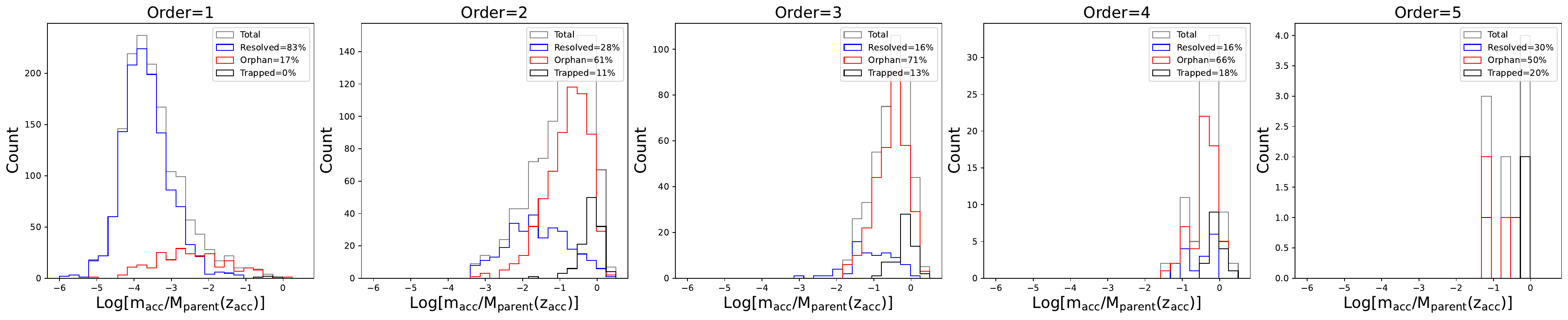}
    \end{subfigure}
    \begin{subfigure}
        \centering
        \includegraphics[scale=0.28]{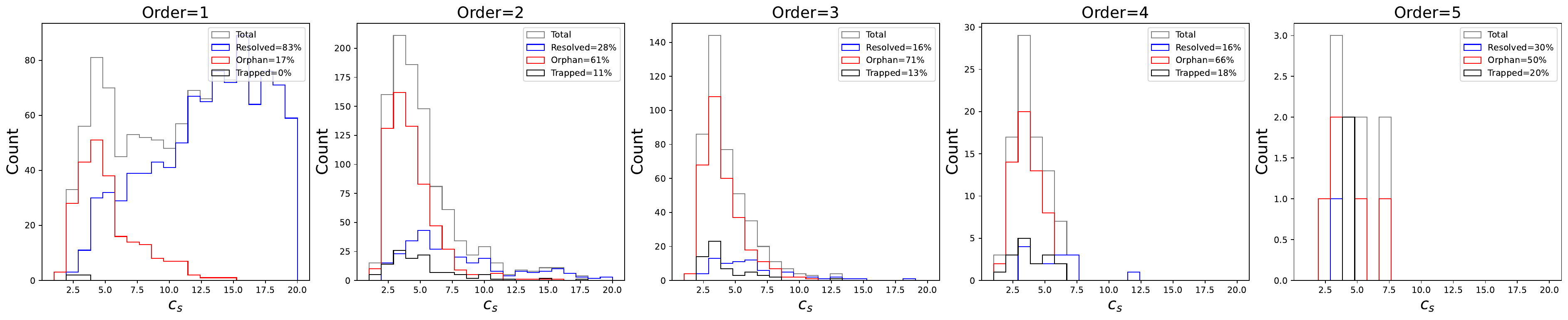}
    \end{subfigure}
    \begin{subfigure}
    \centering
        \includegraphics[scale=0.28]{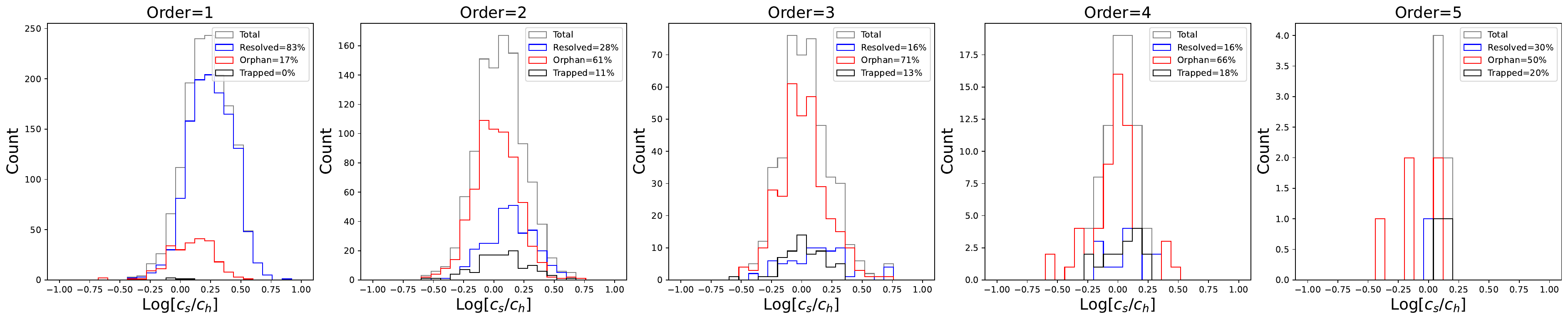}
    \end{subfigure}
    \caption{Distributions of subhalo properties for subhalos of different classes and orders our sample. Each column shows the distribution for subhalos of a given order, while each color shows the distribution for a given class within the order as labeled. From top to bottom the distributions in accretion redshift, merger ratio, subhalo concentration at accretion time and concentration ratio upon accretion are shown respectively.}
    %\zzc{may consider merge order 4 and 5 or remove order 5 (which is not important at all), so making the panels and texts larger for illustration?}}
    \label{fig:order_decompose}
\end{figure*}

To further investigate the influence of subhalo orders, in Fig.~\ref{fig:order_decompose} we show the distribution of the subhalo accretion properties decomposed into different orders as well as different evolution channels. As shown in the top panels, first-order subhalos are mostly accreted at low redshifts, while subhalos accreted at high redshifts are almost all high-order ones. This is because high order subhalos are created by merging low order ones into another halo, such that first order mergers happened at high redshift are converted to high order subhalos at the present day. For a comprehensive analytical model of the redshift and order distribution of halo mergers, we refer the reader to a recent work of Jiang et al. 2024 (in prep).%\citet{Jiang24}. 

The second row of Fig.~\ref{fig:order_decompose} shows the merger ratio distribution of subhalos in different orders. Similar to the distribution of the accretion time, the merger ratio of first-order subhalos is significantly different from that of high-order subhalos. The merger ratios of the former mostly lie in the range of $10^{-5}$ to $10^{-3}$ while the latter are mostly larger than $10^{-2}$. With the order increasing from 2 to 5, the merger ratio distribution tends to concentrate towards 1. This can be understood as the halo mass generally decreases towards the tips of the merger tree, so that the parent mass decreases with the subhalo order. At a given order, the orphan population tend to have larger merger ratios, consistent with Fig.~\ref{fig:massratio_Zacc}.

In the third row of Fig.~\ref{fig:order_decompose}, we show the concentration distribution of the subhalo at its accretion time.\footnote{The concentration for each halo is derived from its radius of maximum circular velocity, $r_{\rm max}$, as provided in the \textsc{hbt+} catalog, assuming $r_{s,0}=r_{\rm{max}}/2.163$ and $c=r_{200}/r_{s,0}$ for an NFW halo.} First-order subhalos accreted at low redshifts tend to have large concentrations, making them resistant to tidal stripping. The orphan population of the first-order subhalos have a lower concentration than the resolved population. For high-order subhalos, most of them have a very low concentration because they are accreted at high redshift before they have been able to build up a more extended envelope. This result is consistent with the findings in \citetalias{Jiang21}. \citetalias{Green21} has found that the concentration ratio between the accreted subhalo and its host halo affects the tidal stripping efficiency when they model this process in SAM. We also check the concentration ratio at accretion time in the bottom row. %\zzc{bette brief the motivation of checking the ratio. is it because Green21 showed that the stripping process depends on the ratio?} 
However, it is hard to segregate the resolved and orphan populations in this parameter space, for every order. 
%\zzc{The ratio for resolved populations looks slightly higher than that for orphans for all orders? but i agree that it is not significnat.}

\begin{table*}
    \centering
    \caption{The fractional contributions of different subhalo classes to the subhalo count at each order, for subhalos with a peak mass between $10^{8}$ and $10^{9}h^{-1}M_{\odot}$ for all six Aquarius halos.}
    \begin{tabular}{l|c|c|c|c|c|r}
    \toprule
         & 1st-order & 2nd-order & 3rd-order & 4th-order & 5th-order & All-order \\
    \midrule
        resolved  & 0.83 & 0.28 & 0.16 & 0.16 & 0.3 & 0.53 \\
    %\midrule
        orphan    & 0.17 & 0.61 & 0.71 & 0.66 & 0.5 & 0.41 \\
    %\midrule
        trapped      & 0    & 0.11 & 0.13 & 0.18 & 0.2 & 0.06 \\
    \midrule
        number of subhalos  & 1513 & 1030 & 451 & 91 & 10 & 3095 \\
    \bottomrule
    \end{tabular}
    \label{tab:order_count}
\end{table*}

To summarize, the results above reveal that the resolved subhalos are mostly first order ones accreted at low redshift through minor mergers and with high concentrations, while orphan subhalos mostly originate from high redshift and high-order accretions which are also dominated by major mergers and low concentrations. %The segregations in subhalo order, merger ratio, concentration and accretion time are closely related to each other, as we expect high redshift accretions are dominated by major mergers with low concentrations and preferentially lead to high order subhalos.

\subsubsection{Physical mechanisms of the subhalo segregation}

%Table.~\ref{tab:order_count} shows the fraction of subhalos belonging to directly accreted and pre-processed modes, and each survival and disruption rate.

 %In \textsc{hbt+}, the accretion time here is defined as the snapshot at which the subhalo was lastly isolated. The parent (sub)halo is the host halo for the given subhalo at the next snapshot, which is not necessarily the final host halo. %Due to our definition, the merger ratio could be larger than 1.
The segregation in subhalo properties can help to explain the different fates of the subhalo populations. We have already discussed that high concentration subhalos are more resistant to tidal stripping and thus more likely to survive. We now further discuss how the segregation in accretion time and merger ratio can lead to different fates of the subhalo populations.

First of all, it is important to realize that the host halo evolves differently at high and low redshifts. It has been well recognized that halo growth consists of two phases, with a fast growth phase at high redshift and a slow growth phase at low redshift~\citep{zhao03,Gao23}. During the fast growth phase, the rapidly growing halo potential can cause the subhalo orbit to decay quickly~\citep{FG84, Ogiya21}, leading to stronger tidal stripping than that in a static potential. This could explain why subhalos accreted at high redshift are mostly orphans.

\begin{figure*}
    \centering
    \begin{subfigure}
        \centering
        \includegraphics[scale=0.5]{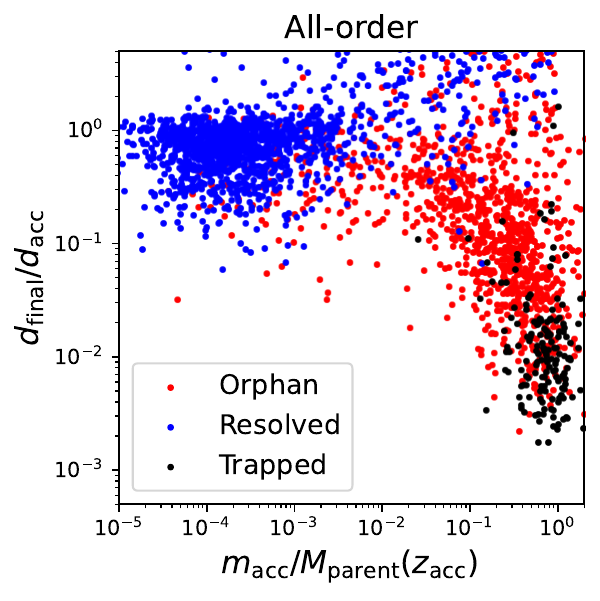}
    \end{subfigure}
    \begin{subfigure}
        \centering
        \includegraphics[scale=0.5]{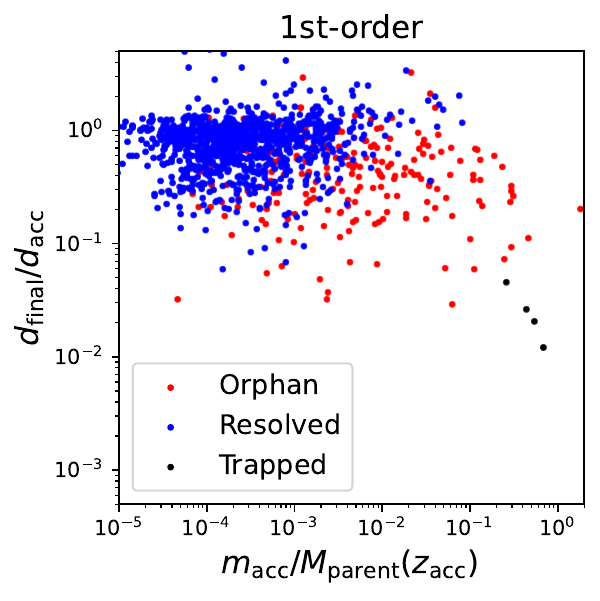}
    \end{subfigure}
    \begin{subfigure}
        \centering
        \includegraphics[scale=0.5]{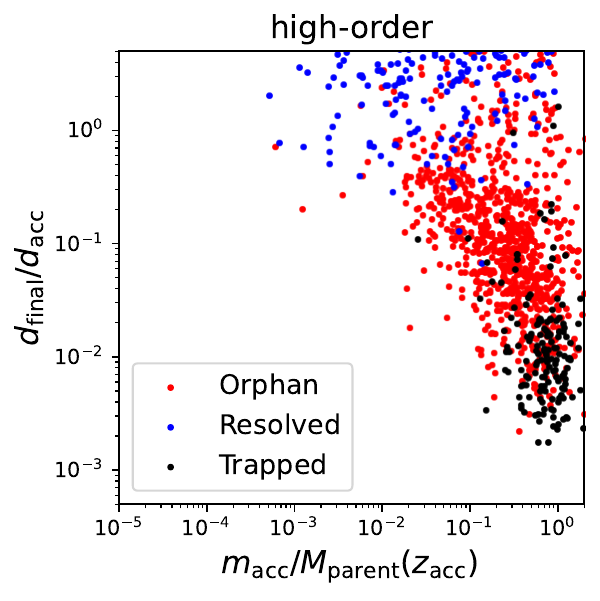}
    \end{subfigure}
    \caption{How the initial merger ratio influences the subhalo orbit decay. The x-axis is the mass ratio between the accreted subhalo and its host/parent halo at accretion time. The y-axis is the host-centric distance of the subhalo at the final time normalized by that at accretion time. The final time is $z=0$ for resolved subhalos and the last resolvable snapshot for orphans. We show the distribution of all subhalos in our sample in the first panel and divide them into first and high order ones in the other two panels.}
    \label{fig:dist_evo}
\end{figure*}

Secondly, the merger ratio is another key parameter in determining the orbital evolution of a subhalo~\citep[e.g.,][]{Jiang08,Jiang14}. Major merger systems typically experience stronger dynamical friction, which quickly drags the subhalo to the inner part of the host halo, leading to stronger tidal stripping. This effect is further boosted by the earlier infall time of the orphan populations, so that the orphans end up closer to the host center than resolved ones at present. This is clearly shown in Fig.~\ref{fig:dist_evo} through the ratio of host-centric distance at the final and initial time. Note the trapped population has sunk to the very center of their hosts by definition. 

Fig.~\ref{fig:dist_evo} also shows that some subhalos (mostly high order ones) run away from their first parent (sub)halos with a distance larger than the initial value. Some reasons could account for this. A subhalo could fly by another (sub)halo at some time and was recorded as a subhalo by \textsc{hbt+}. These fly-by subhalos are more likely to be saved from tidal stripping. On the other hand, the parent (sub)halo also suffers the tidal field from the host halo after it is accreted. If the parent (sub)halo is disrupted, the sub-subhalo belonging to it will be released to other places, again saving the sub-subhalo from the tidal field of its parent (sub)halo. 

%In Fig.~\ref{fig:angularmomentum} we further show the angular momentum evolution of the Aq-A2 subhalos within their parent/host halos. Over a dynamical timescale, the orphan population can lose up to two to three orders of magnitudes of angular momenta, while those of the resolved population only drops by less than one magnitude. Meanwhile, 

% The static host potential model commonly adopted in previous theoretical studies of subhalo evolution can at most apply to the low redshift evolution but not to high redshift where the subhalo orbit is expected to decay due to the rapid host growth~\cite{FG84,SubOrbitDecay}.

\citetalias{vdB18b} studied the disruption of subhalos using a set of idealized simulations. They set their initial merger ratio as $1/1000$, and adopted a static host potential when evolving the subhalos. These conditions are only met by subhalos accreted at low redshift where the host halo grows slowly and the mergers are dominated by minor mergers and first-order accretion. In Fig.~\ref{fig:massratio_Zacc} we draw a black rectangle to indicate these conditions studied in \citetalias{vdB18b}. An interesting finding here is that only one subhalo in this black rectangle is disrupted at $z=0$. It agrees with \citetalias{vdB18b} that subhalos satisfying these physical conditions will survive. The regions outside the black rectangle are not discussed in their \citetalias{vdB18b}, which however is the primary source of orphan subhalos in the simulation. Given the large merger ratio and early accretion time, it is thus reasonable to believe that the extreme mass loss rates found in the simulation for the orphan subhalos are physical.%We will see that, in the right top of Fig.~\ref{fig:massratio_Zacc}, sub-subhalos accreted at high redshift would experience major merger events. The resolved population and orphan population mix here. In addition, trapped subhalos are labelled by dark dots in Fig.~\ref{fig:massratio_Zacc}. They tend to have a merger ratio very close to 1. %and contribute about $6\%$ population in our sample.

\subsection{The extent and influence of artificial disruption}\label{sec:ad}

In the previous section, we analyze the segregation of the resolved and orphan subhalo populations and investigate the origin of the orphan population. However, we still can not separate the withering and artificial disruption channels within the orphan population. \citetalias{vdB18b} argues that subhalos composed of $N$-body particles are inevitably affected by numerical effects. Based on their idealized high-resolution simulations, they propose two criteria to assess the reliability of subhalo evolution in simulations:
%, such as insufficient force softening and discreteness noise caused by the finite number of particles. Based on their idealized high-resolution simulations, they propose two criteria to assess the reliability of subhalo evolution in simulations:

%some brief explanation why they will cause the artificial disruption, like what Symfind did.
The first criterion is related to force softening. \citet{vanKampen2000} advocates that the force softening length to properly resolve the inner structure of a halo should be smaller than the inter-particle separation within the half-mass radius. This means the optimal softening should decrease when a subhalo becomes more stripped. Indeed, \citetalias{vdB18b} showed that a smaller softening tends to make the subhalo survive longer, while insufficient force resolution can lead to increased mass loss. Accordingly, they provide a lower limit of the subhalo bound fraction that can be trusted for a given force softening length, as

\begin{equation}
    f_{\rm{bound}}^{\rm{min1}} = 1.12 \frac{c^{1.26}}{f\left(c\right)^2} \left( \frac{\varepsilon}{r_{s,0}}\right)^2.
    \label{Eq:soft}
\end{equation}
Here, $\varepsilon$ is the softening length of the $N$-body simulation, $c$ and $r_{s,0}$ are the subhalo concentration and scale radius at the initial time, and $f(c) = \rm{ln}(1+c) - c / (1+c)$.

The second criterion arises due to discreteness noise. When a subhalo orbits in the tidal field of the host, subhalo particles that cross the tidal radius get stripped away from the subhalo. In any given time step, the number of particles crossing this tidal boundary varies randomly, following Poisson statistics. In case of a significant random fluctuation, revirialization of the remaining subhalo will cause it to expand more than average, making the subhalo more susceptible to losing even more mass in the next time step. This process can lead to a runaway instability in the mass loss, so that the minimum bound fraction that can be trusted for an initial number of subhalo particles, $N_{\rm peak}$, is given by

\begin{equation}
    f_{\rm{bound}}^{\rm{min2}} = 0.32\left( N_{\rm{peak}}/1000\right)^{-0.8}.
    \label{Eq:Npeak}
\end{equation}
%The two criteria suggest that when the bound fraction $f_{\rm{bound}}=N(t)/N_{\rm{peak}}$ is larger than the criteria, the mass evolution is reliable. When the bound fraction falls below the criteria, mass loss might be attributed to artificial numerical effects rather than physical processes.

To determine if our analysis based on the Aquarius simulation is affected by such numerical effects, we apply these two criteria to our subhalo sample and examine their mass loss curves. %To evaluate Equation~\eqref{Eq:soft}, the scale radius of each progenitor halo is estimated from its maximum circular velocity radius in the catalog, $r_{s,0}=r_{\rm{max}}/2.163$, and the concentration is given by $c=r_{200}/r_{s,0}$.

%We focused on evaluating the bound fraction of subhalos over time and comparing it with the proposed criteria. By doing so, we aimed to identify if and when the subhalos in our sample might have experienced artificial disruption due to numerical effects. %Our goal was to ensure that our conclusions regarding subhalo disruption and survival were not unduly influenced by the limitations of numerical resolution and particle discreteness.

%The questions are whether our discussion based on numerical simulation is affected by numerical effects and whether high-precision numerical simulations like the Aquarius project also have artificial disruption. To answer these questions, we apply the two criteria on our subhalo sample to check the mass loss curve.

\begin{figure*}
    \centering
    \begin{subfigure}
        \centering
        \includegraphics[scale=0.5]{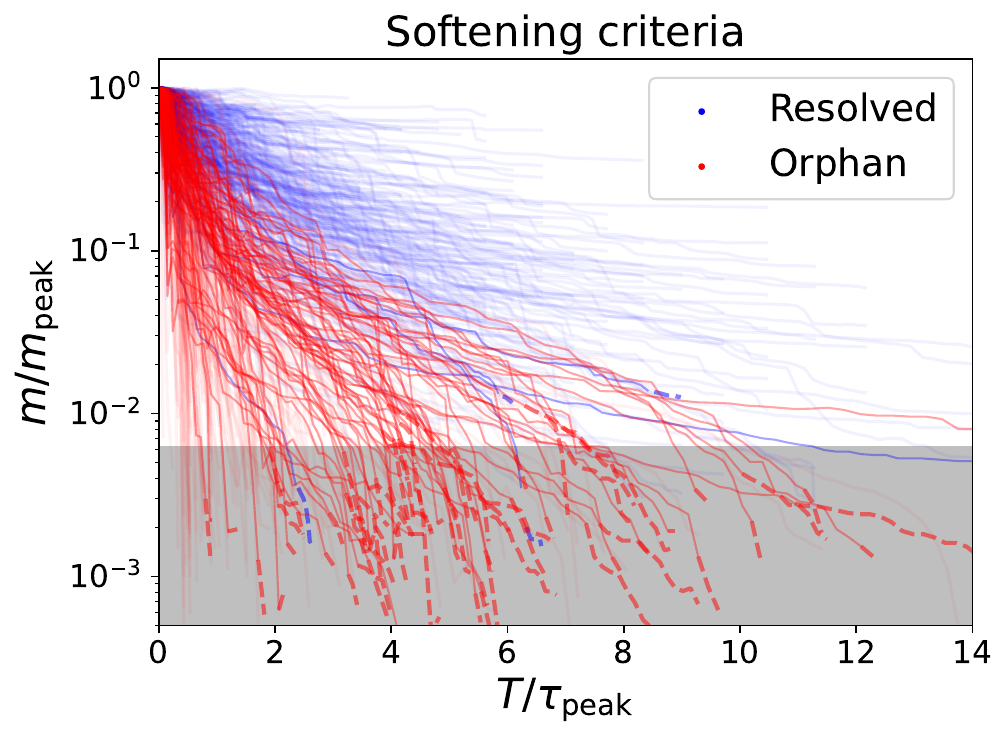}
    \end{subfigure}
    \begin{subfigure}
        \centering
        \includegraphics[scale=0.5]{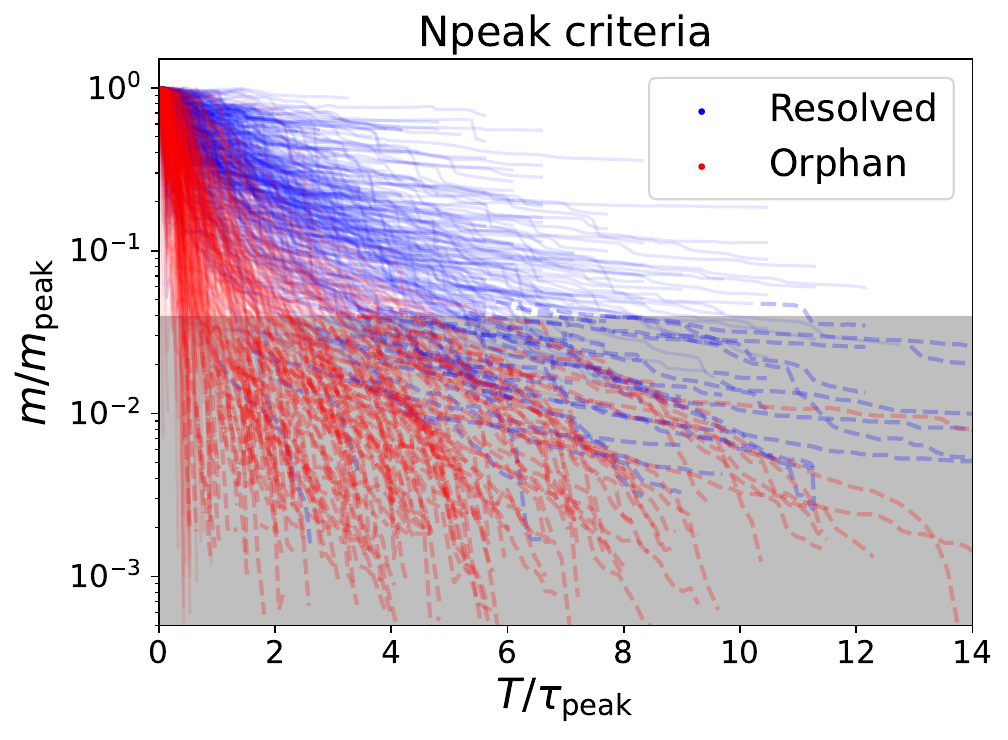}
    \end{subfigure}
    \caption{Same as Fig.~\ref{fig:scaled_massloss}, but showing the effect of the convergence criteria of \citetalias{vdB18b}. For each curve, the part below the softening (Eq.~\ref{Eq:soft}, left panel) or $N_{\rm peak}$ (Eq.~\ref{Eq:Npeak}, right panel) criterion is highlighted as a dashed line. The gray shaded regions mark the range covering 90\% of the starting points of the dashed curves.}%.\jx{only plot the upper 90\% limit, and shade from below.}}
    %\zzc{maybe use a lighter shade, and place shade under the curves? it is hard to see the curves in the shaded region now.}
    \label{fig:vdB_cirteria}
\end{figure*}

The left panel of Fig.~\ref{fig:vdB_cirteria} shows the subhalo mass loss curve scrutinized under the softening criteria for every subhalo in our sample. When a subhalo's bound fraction falls below the threshold given by Eq.~\ref{Eq:soft}, the mass loss curve is shown as a dashed line. It is evident that nearly all resolved subhalos do not fall below the softening bound fraction limit. Only a few orphan subhalos cross this threshold before losing all their masses. These findings suggest that the subhalo mass evolution in our sample is reliable. 

The right panel of Fig.~\ref{fig:vdB_cirteria} dissects the mass loss curves according to the $N_{\rm{peak}}$ criteria. For our subhalo sample with $N_{\rm peak}\sim 10^{4}-10^{5}$, the bound fraction limit is $f_{\rm bound}^{\rm min2}\sim10^{-1.5}-10^{-2}$, which is more stringent than the softening limit. %Since we select the subhalo sample according to their peak mass, the results of Eq.~\ref{Eq:Npeak} should have a similar distribution. 
All of the orphan subhalos cross the bound fraction limit quickly. Even for resolved subhalos, some heavily-stripped subhalos also cross this limit after a few orbital periods.

Through careful analytical modeling and convergence test with the Aquarius simulations, \citetalias{Han16} found that the bound fraction distribution for resolved subhalos follows a log-normal distribution, with a radius-dependent median which is $\sim0.4$ for subhalos located at the virial radius of the host. Although the $N_{\rm{peak}}$ criteria affect the low value tail of the log-normal distribution, the main part of the log-normal distribution remains reliable. In addition, \citetalias{Han16} found that the log-normal part accounts for $\sim 50\%$ of the total accreted subhalo population, while the remaining half, corresponding to orphans here, can be modeled as disrupted ones. According to our analysis, even though we cannot explicitly determine whether the orphan subhalos are disrupted or not due to the numerical effects, we can still classify them as extremely stripped subhalos which can at most add to the very low value tail of the log-normal distribution. Surviving or not, these objects do not introduce significant influences to the statistics of surviving subhalos, and do not invalidate the convergence test of current cosmological simulations. %Thus, while some numerical effects may influence the most stripped subhalos, the primary conclusions regarding the mass distribution of resolved and orphans in the simulation hold true.

%If one believes the statement of \citetalias{vdB18b} that the mass loss of subhalos below the bound fraction limit is affected by numerical effects and is therefore not reliable, then we should focus on the part above this limit for the time being, which is the solid line in Fig.~\ref{fig:vdB_Npeak}. Even after considering the limit of Npeak criteria, there are still significant differences in the mass loss distribution between disrupted subhalos and survival subhalos. These red subhalos often only need two cycles to reach the limit, while blue subhalos take longer to reach or do not experience possible discreteness noise due to slow mass loss. Therefore, we can believe that at least the final bound fraction of these disrupted subhalos will definitely fall below this limit, which will not affect the statistical distribution of larger bound fractions. The reasons for the rapid mass loss of these subhalos have also been analyzed in the previous section.

\citetalias{vdB18b} expressed concern that performing convergence tests solely by increasing the number of particles, without considering the convergence of softening, could lead to false convergence. By combining the two criteria in our examination, we assert that current state-of-the-art cosmological simulations should be free from the insufficient softening length problem. Although the bound fraction in simulations still suffers from the limitations of discrete noise, this issue is related only to the number of particles. Thus, we can rely on the traditional convergence test at this time. The Aquarius simulation, with its five sets of resolution runs from low to high, demonstrates that the subhalo mass function is indeed convergent\citep{Aquarius, Han16}. Given that the subhalo statistics in the Aquarius simulations are also consistent with well established results from other cosmological simulations~\citep[e.g.,][]{Aquarius,Han16,HBT+}, these results imply that artificial disruption caused by discrete noise or insufficient softening will not significantly affect our current understanding of the subhalo population.

Finally, we comment that the two criteria given by \citetalias{vdB18b} are based on their own simulation pipelines which could have different performance from others. For example, \citet{Aoife} tried to reproduce the results of \citetalias{vdB18b} using both the \textsc{Gadget-4} code and the \textsc{treecode} used in \citetalias{vdB18b}. They also tested the postprocessing using the original code of \citetalias{vdB18b} and \textsc{subfind}. The results show that subhalos in the \textsc{Gadget-4} simulation are slightly more resilient than those in the \textsc{treecode}, and there is also a systematic offset in the subhalo mass identified by the two subhalo finders. Therefore, the exact convergence criteria could vary for different simulation codes and analysis pipelines (see also Appendix~\ref{sec:convergence}).

While numerical effects cannot be entirely ruled out, the core findings about the differing distributions between orphan and resolved subhalos provide important insights into the physical processes. Furthermore, our use of \textsc{hbt+} ensures robust and continuous tracking of subhalo evolution, reinforcing the reliability of our results within the established numerical limits \citep[see also][]{Mansfield23}.

\section{The fate of disrupted subhalos in Idealized Simulation}\label{sec:idealsim}
As discussed above, the numerical experiments in \citetalias{vdB18b} do not match the typical initial conditions of orphan subhalos found in cosmological simulations. To verify that our conclusions are robust to numerical effects, in this section we investigate the evolution of a typical subhalo generated from a high redshift major merger using idealized simulations.

\subsection{Initial Conditions}
 %We should make sure the initial conditions match the situation of the real cosmology context.
We set the subhalo initial mass as $m=10^{8}h^{-1}M_{\odot}$ and the host halo virial mass as $M_h=10^{9}h^{-1}M_{\odot}$. Both of them have a halo concentration of $5$ at an initial redshift of 4. The subhalo is released from the virial radius of the host at an initial velocity of $0.5V_h$ with a pure tangential component, where $V_h=\sqrt{GM_h/R_h}$ is the virial velocity of the host.%The initial subhalo is ejected at the location where its edge is tangent to the edge of the host. The boundary of the halo is $r_{200}$ here. We only set the initial tangential velocity component and no radial velocity component. 
%The velocity is half of $V_h$, where $V_{h}=\sqrt{GM_h/R_h}$. 

We sample both the host and the subhalo with $N$-body particles, so that dynamical friction can be resolved. The initial particle distribution of each halo is constructed using \textsc{AGAMA}~\citep{agama}, which builds a steady-state phase space distribution function for the adopted halo mass profile through Eddington inversion. We adopt an NFW profile with a squared-exponential truncation beyond virial radius following \citet{2023MNRAS.518.5356L}. The truncation is introduced to avoid the divergence in the integrated mass of the NFW profile. 

\subsection{Simulation}

With the initial conditions above, we first run a high resolution simulation using $10^{7}$ particles for both the satellite and the host halo, with particle masses of $m_{p,s}=10h^{-1}M_{\odot}$ and $m_{p,h}=10^{2}h^{-1}M_{\odot}$ for the satellite and the host respectively. %It is a compromise to the computational cost. It could cause some artificial effects but we believe the influence is weak. %need citation here
The force softening length is $\varepsilon=0.003r_{s,s}$, where $r_{s,s}$ is the scale radius of the subhalo, corresponding to the highest force resolution used in \citetalias{vdB18b}. 
The system is simulated with the \textsc{Gadget-4} code~\citep{gadget4} for 10Gyr. We use \textsc{hbt+} to identify the bound structure of the subhalo during its evolution.

According to Equation~\eqref{Eq:soft}, the adopted softening length can reliably resolve the bound fraction down to $10^{-4}$, while Equation~\eqref{Eq:Npeak} also predicts a comparable limit. This places a conservative limit on the reliable range of the mass loss curve. As we show in Appendix~\ref{sec:convergence}, our simulations are expected to be reliable down to a slightly lower bound fraction limit.

\begin{figure*}
    \centering  
    \includegraphics[scale=0.4]{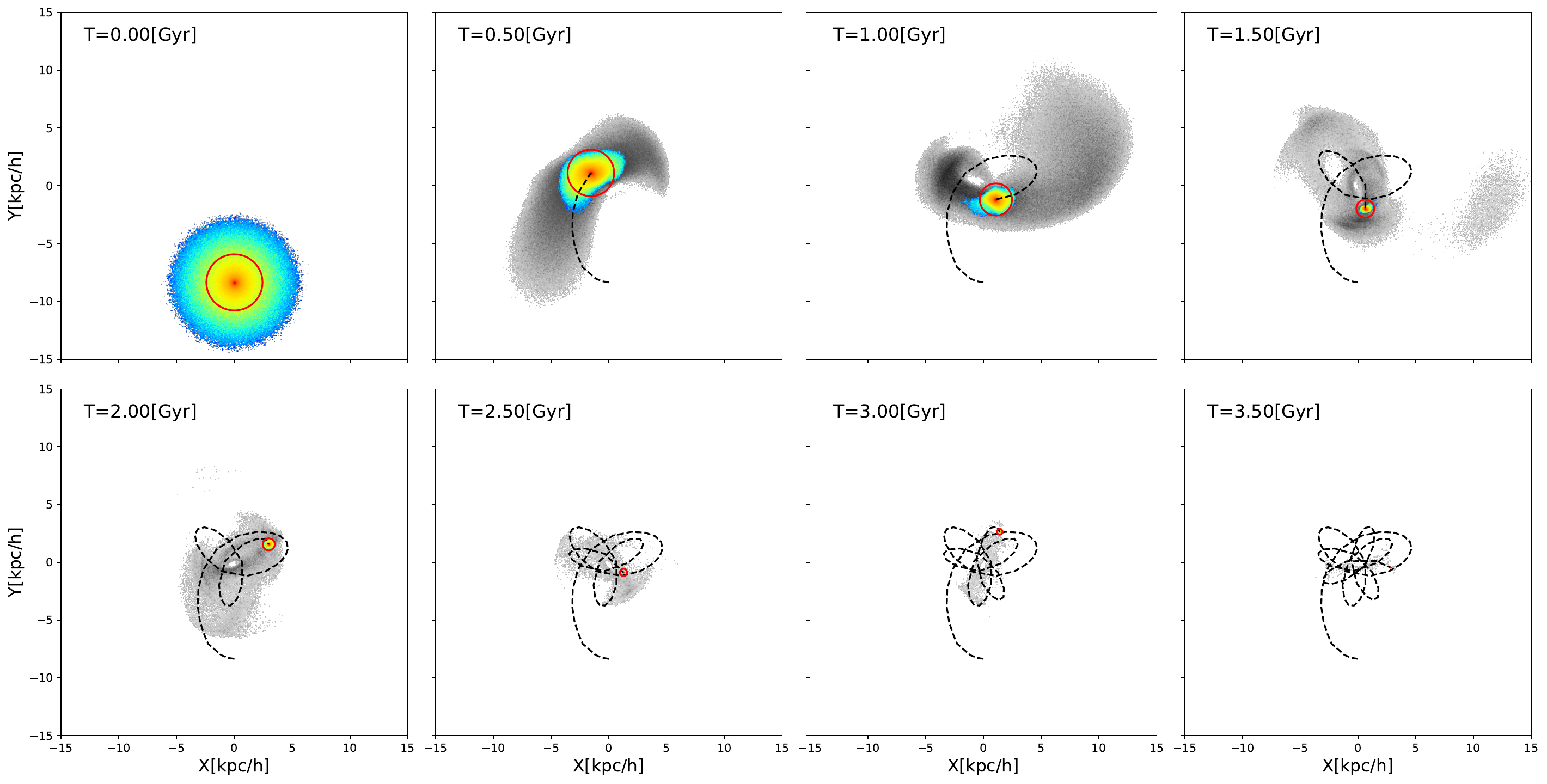}
    \caption{The snapshots of our idealized simulation. The red circle shows the $r_{200}$ of the bound part of the subhalo, and the colored dots inside it show the bound particles, with redder color indicating higher density. The gray dots outside it show the unbound particles. The black dashed line shows the trajectory of the subhalo in the host halo. The orbit shrinks into the inner region of the host halo due to dynamical friction, and the subhalo becomes unresolved after $3.5\rm{Gyr}$. We do not plot the host halo particles here.}
    \label{fig:fiducialsnapshot}
\end{figure*}

Fig.~\ref{fig:fiducialsnapshot} shows several snapshots of the simulation.   %The blue particles represent the bound particles and the red particles represent the unbound particles which are stripped away by the tidal force. It is noted that we do not plot the host halo particles in Fig.~\ref{fig:fiducialsnapshot}. 
%In our simulation set, the host potential is modelled with $N$-body particles so that dynamical friction effects will play an important role. 
As shown by the gray dashed line, the trajectory of the subhalo shrinks to the inner region of the host after a few orbital periods. %The enhanced tidal effect strips the material of the subhalo. 
During this process, the stripped particles form several shells which also expand over time as the subhalo becomes more disrupted and its self-gravity weakens. After about 3.5Gyr, the self-bound subhalo can no longer be resolved at the current resolution. 

\begin{figure*}
    \centering
    \includegraphics[scale=0.5]{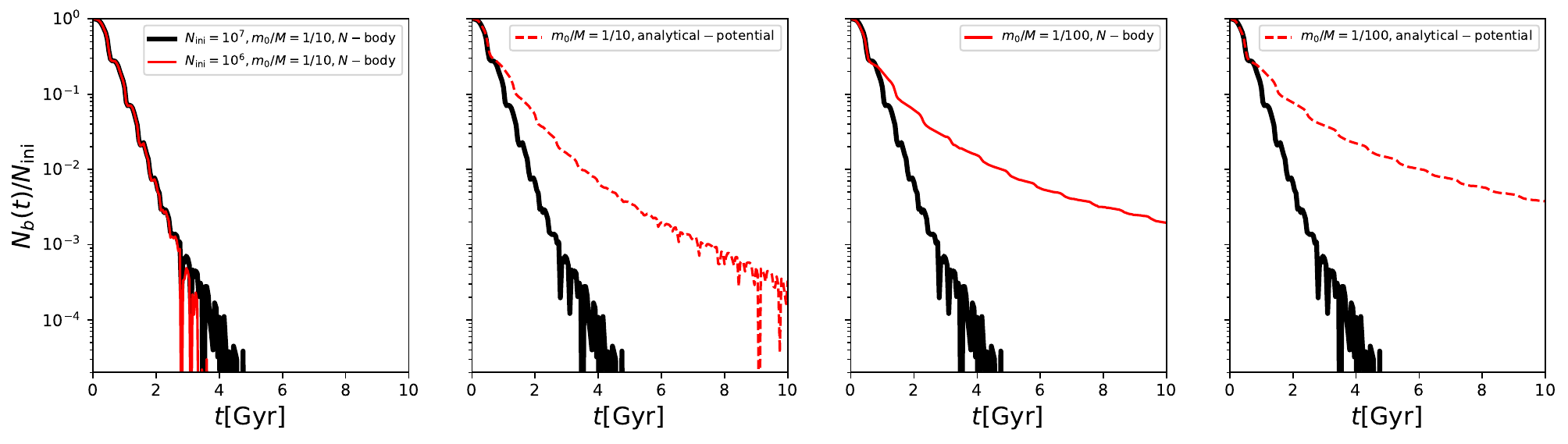}
    \caption{The mass loss curves of our idealized simulations. The black line in each panel represents the high resolution run, which resolves both the satellite and the host halo with an initial particle number $N_{\rm{ini}}=10^{7}$. The red curves in the four panels show the same system evolved using different setups for the satellite or the host relative to the high resolution run. In the first panel, the subhalo is resolved using only $N_{\rm{ini}}=10^{6}$ particles. In the second panel, the host halo is replaced with an analytical potential. In the third panel, the initial mass of the subhalo is reduced by a factor of 10. In the fourth panel, both a $1/10$ initial subhalo mass and an analytical host potential are adopted. For all red curves, the subhalo has an initial particle number $N_{\rm{ini}}=10^{6}$.}
    \label{fig:fiducialmassloss}
\end{figure*}

The mass evolution of the subhalo is plotted in the first panel of Fig.~\ref{fig:fiducialmassloss}, which shows an exponential mass loss at a rate of about $90\%$ per Gyr. At this extremely large mass loss rate, the subhalo would be stripped to less than $10^{-10}$ of its initial mass if evolved to $z=0$. This matches what we discussed in the last section, that the pre-processed subhalo undergoing major events will lose almost all of its mass in $1\sim2$ Gyr. Meanwhile, as shown by the \citetalias{Han16} model and the discussion in section~\ref{sec:property_segregation}, most of the resolved subhalos have a bound fraction larger than $10^{-3}$.

As we have discussed, dynamical friction is an important factor contributing to the high mass rate for these major merger systems. To demonstrate this, in the second panel of Fig.~\ref{fig:fiducialmassloss} we show the mass evolution of the same subhalo if we evolve it in an analytical NFW potential rather than the $N$-body one. As there is no density wake in the analytically-modeled host halo, dynamical friction is absent in this setup. However, there will still be self-friction from the drag of stripped particles, so that the orbit of the subhalo will still shrink over time~\citep{Miller20}. Despite this, it is evident that the mass loss is now much slower without dynamical friction, due to the slower decay of the subhalo orbit.

To reduce the computational cost, we have sampled the subhalo with 10 times less particles than in the high resolution run. To ensure that the mass evolution is reliable, in the left panel we also show the mass evolution of a subhalo resolved with $10^{6}$ particles in the full $N$-body run, while keeping $10^7$ particles for the host halo. The mass evolution in the low resolution run only starts to degrade at a bound fraction less than $10^{-3}$, ensuring that the low resolution run is sufficient above this mass fraction. 

%The red dashed line in the second panel shows the case if the subhalo evolves in a static analytical potential rather than an $N$-body host system. All other sets are the same as the fiducial simulation which is shown by the black solid line same as that in the first panel. The difference leads to the disappearance of the dynamical friction effect, only self-friction exists. The self-friction means that the unbound particles stripped by the tidal field will also drag the motion of the bound structure, though the effect is unignorable it is still weaker than that of dynamical friction. Before the first peri-centric passage, the mass loss rate is similar to the fiducial one. After that, the mass loss rate is dramatically suppressed. At 10Gyr, a typical time scale of the subhalo evolution, the final bound fraction is about $10^{-4}$. 

%In the following, we run the simulation to compare the divergence of the mass loss rate.

In order for dynamical friction to be important, it is essential to have a large merger mass ratio. The third panel shows the case if the subhalo is accreted through a minor merger of $1:100$ ratio, rather than $1:10$. The particle masses of both halos are the same as in the original high resolution run. In the right panel, we further suppress dynamical friction by evolving the subhalo in an analytical potential. For this minor merger system, the mass loss is much slower than the subhalo in the original run, due to the slow evolution of the orbit. In addition, the $N$-body run and the analytical potential run now produce very similar results, further supporting that dynamical friction is unimportant here. 

We note that the idealized simulation run by \citetalias{vdB18b} is very close to the last case of our simulations above. They studied minor mergers and found that the subhalo will never disrupt in the highest resolution. We illustrate that the properties of accreted subhalos will significantly affect the mass loss rate. Most of the disruption (orphan) we detected in the simulation was just caused by the extreme mass loss rate that is not covered in their study. %The mechanism of the subhalo disruption should still be assigned to the \textit{Withering} channel suggested by \citet{vdB17}. But we show here that the above case is very common. %We think that artificial disruption is very common in cosmological simulation but they do not affect the \textit{Withering} of the subhalo in nature. 
%As shown in the first panel of Fig.~\ref{fig:fiducialmassloss}, the subhalo run in the median resolution disappears at 3Gyr, while increasing the resolution to the highest level will not change their withering fate. 

Our results are also compatible with the findings in \citet{Errani21}. They define a structure parameter $T_{\rm{max}}=r_{\rm{max}}$/$V_{\rm{max}}$ of the subhalo, and claim that the evolution of $T_{\rm{max}}$ (named ``tidal-track'') has an asymptotic endpoint if a subhalo has a cuspy density profile. The endpoint of the tidal track is determined by $T_{\rm{peri}}$ which is the period of a circular orbit at the pericenter of the subhalo. However, the conclusion of \citet{Errani21} is based on the assumption that the orbit of the subhalo will not change. For subhalos experiencing strong dynamical and self frictions, the decrease in dynamical time keeps lowering the endpoint of the tidal track and thus a fast mass loss can be maintained.

\section{The statistical contribution of subhalo disruption with a semi-analytical model}\label{sec:satgen}

In the previous sections, we have shown that the orphan and resolved populations are distinct in their mass loss rates. Despite this, numerical effects still hamper us from reliably resolving the final mass of a subhalo at the lowest mass end for any finite mass resolution. To assess the influence of numerical effects qualitatively, we can evolve the subhalos according to an analytical model and compare the results with the simulation.

To this end, we apply the semi-analytical model (SAM) {\tt\string SatGen} \citep{Jiang21,Green21} to the subhalo populations in our simulation. {\tt\string SatGen} follows the evolution of subhalos and satellites through a series of physical recipes, including calibrated descriptions of dynamical friction and tidal stripping. 
% While {\tt\string SatGen} can start with subhalo merger trees generated by MCMC realization of the EPS theory, here instead we apply it to the merger tree extracted by \textsc{hbt+} in the Aquarius simulation.
While {\tt\string SatGen} can start with subhalos generated by Monte Carlo merger trees using the EPS theory and evolve their orbits by orbital integration with dynamical friction, here instead we use the merger trees extracted by \textsc{hbt+} and actual subhalo orbits in the Aquarius simulation.
%Some free parameters are needed in the above functions, they should be calibrated by the simulation. %To avoid the influence of the numerical effects from the cosmological simulation, the authors calibrated their model parameters by the DASH library \citep{Ogiya19}, which is composed of a large number of idealized simulations.

%I will briefly introduce the method of the {\tt\string SatGen} dealing with the non-linear evolution of the subhalo. 
To model tidal stripping, \satgen assumes that at each time step, the stripped mass is proportional to the subhalo mass outside its instantaneous tidal radius\citep[e.g.][]{Zentner05, Pullen14},
\begin{equation}
    \Delta m=-\alpha \frac{\Delta t}{t_{\text {char }}} m\left(>l_{\mathrm{t}}\right) .
    \label{eq:tidal_massloss}
\end{equation}
Here, $t_{\rm{char}}$ is the characteristic orbital time of the subhalo, and $l_t$ is the tidal radius defined following \citet{Klypin99}. $\alpha$ is a fudge factor describing the efficiency of tidal stripping, calibrated with the DASH library of subhalo evolution in a large number of idealized simulations~\citep{Ogiya19} by \citet{Green21}. They find that $\alpha$ has a typical value of 0.55 and a dependence on the concentration ratio between the subhalo and its host halo.

To follow the mass evolution of a given subhalo, {\tt\string SatGen} takes several parameters as input, including the mass and concentration of the subhalo at its accretion time, the instantaneous mass and concentration of its host halo, and the orbit of the subhalo. When dealing with high order subhalos, we evaluate tidal stripping from both the parent subhalo and the host halo separately, and take the larger of the two to account for the competing stripping from them.

The original \satgen code includes a recipe for subhalo disruption, which happens when the radius of a subhalo has been stripped below a critical radius of 0.77$r_s$ according to \citet{Hayashi03}. However, as CDM subhalos are robust to disruption~\citep{vdB18a, Errani21}, we switch off this implementation in our application. %\zzc{as we switched it off, why bother introducing the details? I believe by default it is not enabled, because AD is something people just want to get rid of. or we can say "The \satgen code also includes an optional recipe to mimic subhalo artificial disruption, which we switched off here." (is this description correct?)}\fhc{Green21 compared the results obtained with and without applying the disruption recipe. In principle, this optional recipe can not reflect the nature of artificial disruption, according to vdB18a. Maybe "mimic" is not a suitable description here.}
There are also cases when \satgen fails to evaluate the tidal radius of a subhalo, leading to a sudden mass drop to below the mass resolution of the code. We correct for such failures by assigning a mass of 1/10 the subhalo mass at the previous step, preventing a subhalo from complete disruption in our analysis.

In Appendix~\ref{sec:satgen_example} we show two examples of the mass and density evolution predicted by \satgen. Overall, the model evolution agrees with the mass evolution in the simulation.

\subsection{Results from \satgen}\label{sec:satgen_result}

\begin{figure*}
    \centering
    \begin{subfigure}
        \centering
        \includegraphics[scale=0.7]{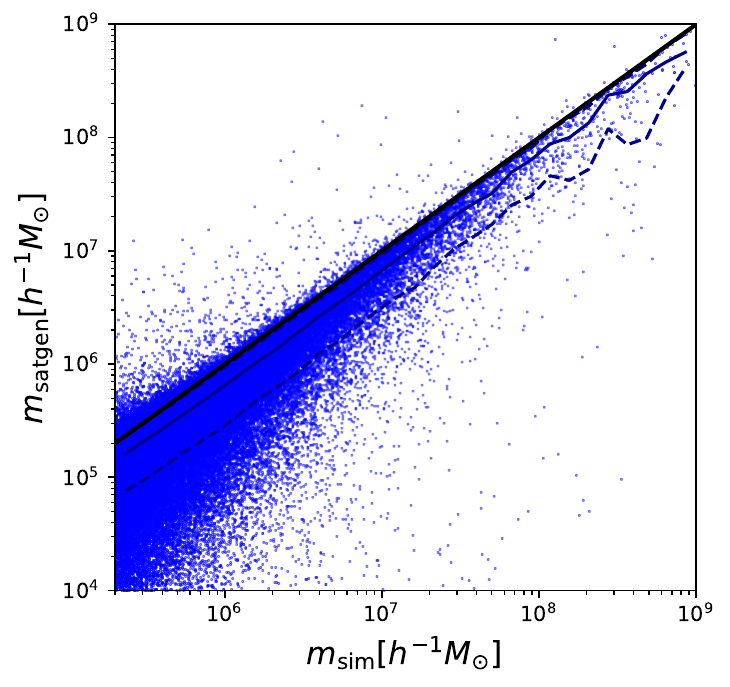}
    \end{subfigure}
    \begin{subfigure}
        \centering
        \includegraphics[scale=0.7]{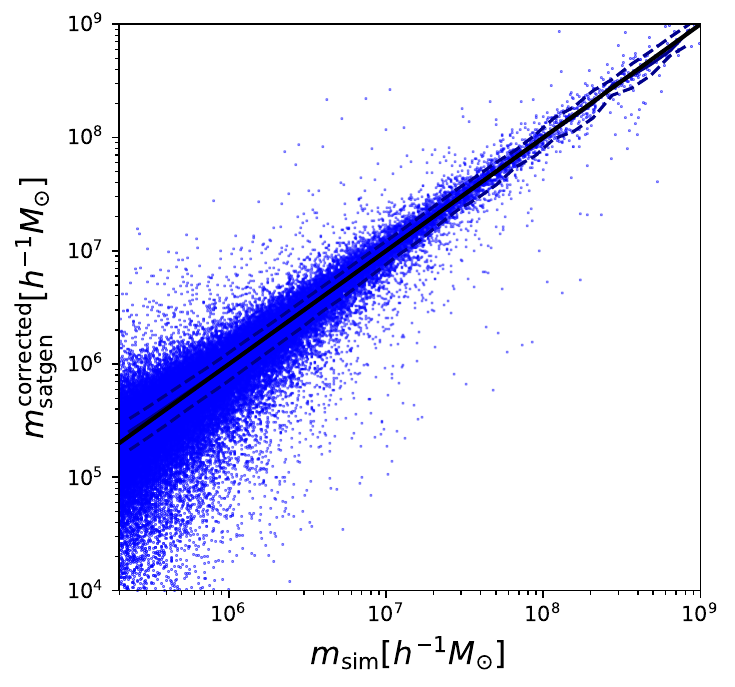}
    \end{subfigure}
    \caption{The subhalo masses in the simulation compared with the {\tt\string SatGen} predictions. Each blue dot represents a resolved subhalo in simulation at $z=0$. The blue solid and the dashed lines show the median, and the $16\%$ and $84\%$ percentiles at a given simulated mass, respectively. The black solid line indicates the 1-to-1 ratio for reference. The left panel shows the original \satgen prediction while the right panel shows the result after adding a mass correction to the prediction (see text for details).
    }
    \label{fig:satgen_mass}
\end{figure*}

Fig.~\ref{fig:satgen_mass} compares the final subhalo masses predicted by {\tt\string SatGen} with those from the simulation for the resolved subhalos. On average, the simulated mass tends to be underpredicted by \satgen by $\sim40\%$. %The result indicates the {\tt\string SatGen} model supports that the mass loss evolution in the $N$-body simulation should be reliable for the resolved population. What's more, the {\tt\string SatGen} mass underestimates the real subhalo mass about $40\%$ statistically at $z=0$.  
This difference is probably because {\tt\string SatGen} calibrated the parameter $\alpha$ in Eq.~\ref{eq:tidal_massloss} with DASH which used the unbinding algorithm by \citetalias{vdB18b} to track subhalos, different from \textsc{hbt+} used Aquarius. It has also already been found that the algorithm of \citetalias{vdB18b} tends to underestimate the subhalo mass compared to conventional subhalo finders like \textsc{subfind}, consistent with our findings here~\citep{Aoife}. Note the subhalo masses returned by \textsc{hbt+} are generally even larger than \textsc{subfind}~\citep{Han12a,HBT+}. %We have discussed the difference in sec.~\ref{sec:ad} that they will underestimate the subhalo mass compared to the normal unbinding algorithm adopted in cosmological simulations.
%\zzc{I disagree with reason 2. DASH also uses bound mass (despite the difference in finding algorithm), so SatGen is calibrated with bound mass rather than tidal mass. }

%To calibrate the mass difference, we fit the median distribution in Fig.~\ref{fig:satgen_mass}, which could be taken as the ratio between $m_{\rm{satgen}}$ and $m_{\rm{sim}}$. The formula is
In principle, we can recalibrate the $\alpha$ parameter in {\tt\string SatGen} using our simulation, which however is computationally expensive. Instead, we choose to rescale the modeled mass loss through
%Here we adopt an effective correction for simplicity. To have a more detailed discussion on the evolution of the subhalo, the better way is to investigate the bound fraction($\mu=m/m_{\rm{peak}}$) of each subhalo. Fig.~\ref{fig:satgen_mu} shows the bound fraction distribution of the Aquarius subhalos for each $m_{\rm{peak}}$ bin. The $\alpha$ in Eq.~\ref{eq:tidal_massloss} represents the mass loss efficiency. The longer the evolution time, the more mass will be stripped, which leads to a larger mass bias for the subhalo with a lower bound fraction. We calibrate the mass bias by a simple formula:
\begin{equation}
m_{\rm{satgen}}^{\rm{corrected}}/m_{\rm peak}=(m_{\rm{satgen}}/m_{\rm{peak}})^{0.65}.
\end{equation}
This is approximately equivalent to rescaling the $\alpha$ parameter to $0.65\alpha$ in the model.
%\zzc{This is approximately equivalent to use $1/1.3\alpha$ instead of $\alpha$ in Eq.~\ref{eq:tidal_massloss}.} 
%By doing so, The corrected mass $m_{\rm{satgen}^{\rm{corrected}}}$ will have no underestimate of the subhalo mass from our simulation, 
As shown in the right panel of Fig.~\ref{fig:satgen_mass}, the median distribution of the \emph{bias-corrected} {\tt\string SatGen} prediction now agrees with the simulation.% could lie on the diagonal lines of diagrams of Fig.~\ref{fig:satgen_mu}.    %We compare the bound fraction calculated by {\tt\string SatGen} with that from the simulation, both resolved subhalos and orphan subhalos. 

\begin{figure*}
    \centering
    \begin{subfigure}
        \centering
        \includegraphics[scale=0.45]{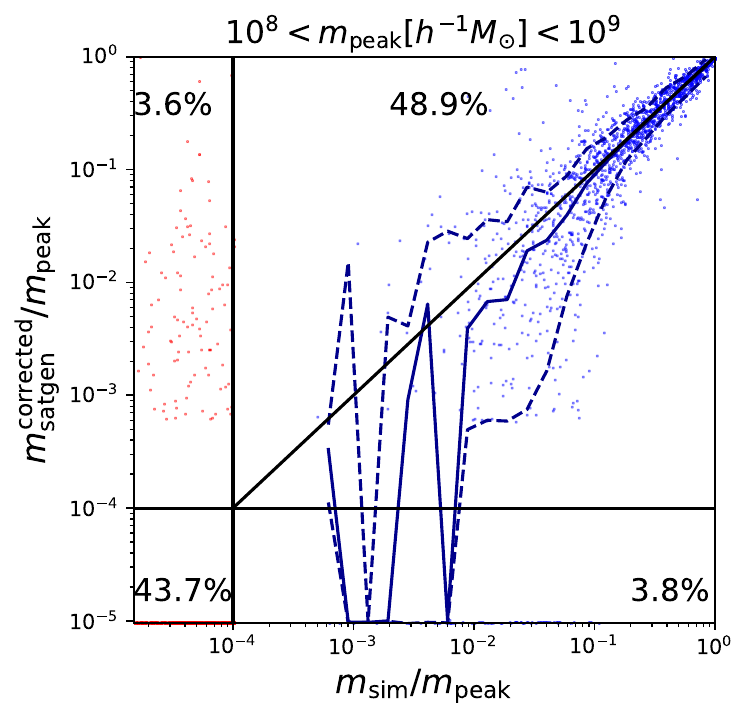}
    \end{subfigure}
    \begin{subfigure}
        \centering
        \includegraphics[scale=0.45]{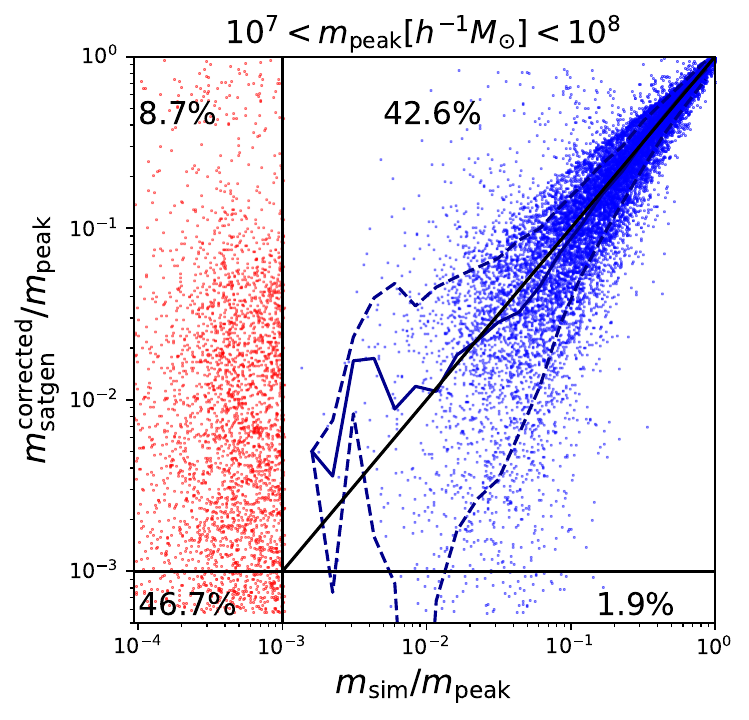}
    \end{subfigure}
    \begin{subfigure}
        \centering
        \includegraphics[scale=0.45]{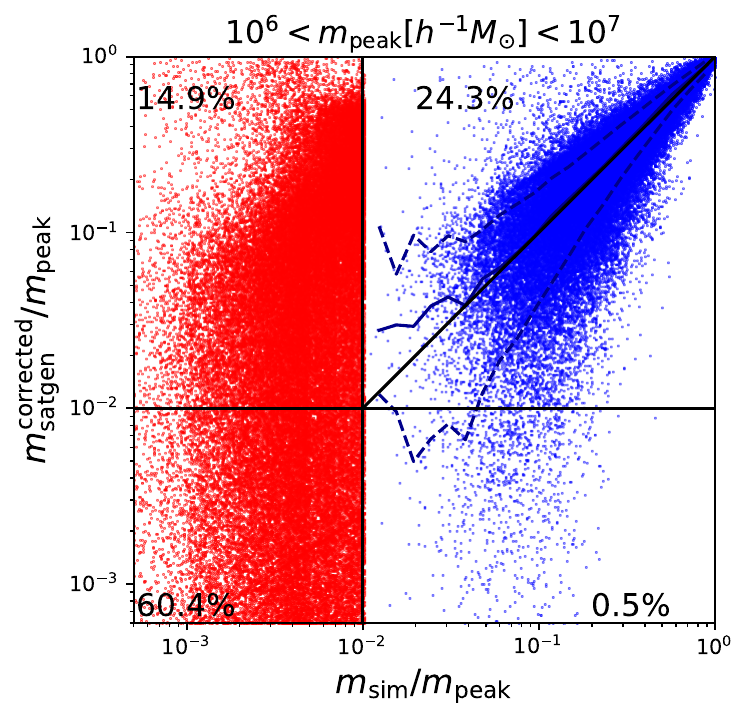}
    \end{subfigure}
    \caption{The bound fraction predicted by {\tt\string SatGen} (bias-corrected) compared with that in the simulation for all the subhalos. The three panels show the subhalos in different $m_{\rm{peak}}$ bins as labeled. The diagonal solid line is the 1-to-1 ratio curve, and the blue curves show the median and the 16th and 84th percentiles at a given simulated mass. The horizontal and vertical curves mark the boundary of orphan subhalos resolved with no more than one particle in the corresponding model. The number in each quadrant shows the fraction of subhalos residing in the quadrant.
    }
    \label{fig:satgen_mu}
\end{figure*}

To investigate the \satgen predictions on the orphan subhalos, Fig.~\ref{fig:satgen_mu} further compares the predicted and simulation bound fractions for each $m_{\rm{peak}}$ bin. Each panel of Fig.~\ref{fig:satgen_mu} is divided into four quadrants 
by the resolution limit $\mu=m_{\rm{p}} / \rm{MIN}(m_{peak})$,
% . The boundary of the quadrants is calculated at $\mu=m_{\rm{p}} / \rm{MIN}(m_{peak})$, 
where $m_{\rm{p}}$ is the particle mass of the simulation and $\rm{MIN}(m_{peak})$ is the lower bound of the peak-mass bin, so that subhalos with less than one particle all lie below this critical ratio. 
%\zzc{the reverse? subhalos with less than one particle all lie below this critical ratio}
%\zzc{one particle or 20 particles? (we don't have 5-particle subhalo in HBT, right?) should we use the HBT subhalo limit instead of single particle mass?}\fhc{The true bound fraction of the unresolved subhalo is unknown. The maximum bound fraction of unresolved subhalos is 1/MIN(Npeak) and the minimum bf of resolved subhalos is 20/MAX(Npeak). Some overlap could be introduced if using 20 particles. But the true distribution of bound fractions between 1-20 particles should indeed be continuous.}

Subhalos in Quadrant 1 (top right) are predicted to be resolved ones in both simulation and \satgen, and those in Quadrant 3 (bottom left) are consistently predicted to be unresolved. Quadrant 2 (top left) contains subhalos unresolved in simulation but predicted to be resolved by \satgen, thus likely experiencing artificial disruption. Quadrant 4 (bottom right) contains resolved subhalos that are predicted unresolved by \satgen, representing catastrophic failure in the \satgen prediction.

For the left panel of Fig.~\ref{fig:satgen_mu}, subhalos in this panel have about $10^{4} \sim 10^{5}$ particles at their accretion time. $43.7\%$ of the subhalos fall in Quadrant 3 while $3.6\%$ of the subhalos fall in the artificial disruption region. With the decrease of the subhalo maximum particle number, as shown by the middle and right panels, the percentage of the subhalos in quadrant two increases. Especially in the third panel, the fraction in Quadrant 2 grows up to about $15\%$. It suggests that artificial disruption becomes a bigger concern at lower $N_{\rm{peak}}$. The result agrees with the discussion in sec.~\ref{sec:ad}, that the discreteness noise affects the subhalo population with an insufficient particle number. Fortunately, considering that the subhaloes in the right panel have merely hundreds of particles at their accretion time, only the subhalos of dozens of particles in $z=0$ are affected by the artificial disruption.
%Fortunately, subhalos in the right panel have hundreds of particles at their accretion time, even though the artificial disruption exists here, the numerical effects will only pollute subhalos with dozens of particles at $z=0$.

\begin{figure}
    \centering
    \includegraphics[width=\columnwidth]{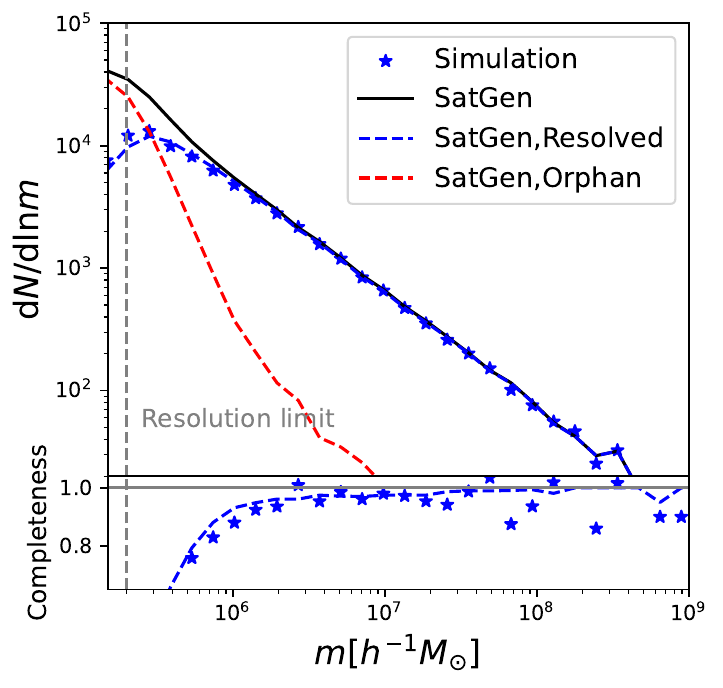}
    \caption{The subhalo mass function predicted by {\tt\string SatGen}. The blue star shows the subhalo mass function resolved in the Aquarius simulation. The blue dashed line shows the \emph{bias-corrected} {\tt\string SatGen} prediction on the resolved population, while the red dashed line shows its prediction on the orphan subhalos. The black solid line is the total subhalo mass function predicted by {\tt\string SatGen}, which can be regarded as a theoretical subhalo mass function without artificial disruption. The ratios between the resolved mass functions and the total \satgen prediction are shown in the bottom panel. The vertical dashed line marks the subhalo mass resolution (20 particles) of the simulation.}
    \label{fig:satgen_massfunction}
\end{figure}

Fig.~\ref{fig:satgen_massfunction} shows the final subhalo mass function from the Aquarius simulation and the {\tt\string SatGen} model with the mass correction. It is not surprising that the subhalo mass function predicted by {\tt\string SatGen} for resolved subhalos is very similar to that of the simulation. For the abundance of the orphan population predicted by {\tt\string SatGen}, the part inside the resolution limit corresponds to Quadrant 3 in Fig.~\ref{fig:satgen_mu}, while that above the limit is contributed by Quadrant 2 which can be attributed to artificial disruption. This red dashed line thus quantifies the influence of artificial disruption on the subhalo mass distribution. 
It dominates the region of $m_{\rm{sub}} < 10^{6}h^{-1}M_{\odot}$ (about 100 particles), where the $N$-body simulation underestimate the model mass function by more than $10\%$. With the increase of the subhalo mass (particle number), the impact of artificial disruption attenuates quickly. For subhalo mass larger than $3\times 10^{6} h^{-1}M_{\odot}$ (about 300 particles), the impact is less than $3\%$. We can thus conclude that the artificial disruption is not a concern for the subhalo population with enough particle numbers.
 
\section{Discussion: disruption of the stellar component and orphan galaxies}\label{sec:discussion}

%\subsection{total disruption rate and unevolved subhalo mass function for different orders}

%\subsection{Stellar mass stripping rate and orphan galaxies}

%Some introducation to Orphan galaxies

%\textit{Orphan galaxy} refers to the satellite which has lost a significant portion of its dark matter subhalos due to tidal effects from the host potential or other mechanisms. More simplified, 
When a subhalo becomes unresolved in the simulation, the satellite associated with the subhalo is commonly modeled as an orphan galaxy. Orphan galaxies are invoked in many semi-analytical or empirical models to bridge the gap between theoretical predictions and observational constraints~\citep[e.g.,][]{Springel01, Gao04L, Guo11, Campbell18, UM, DeRose22}. However, some recent constraints from gravitational lensing indicate that the empirical model overestimates the abundance of the orphan galaxies~\citep{Kumar24}. \citet{Bahe19, Sifon23} also found that the satellite embedded within the disrupted subhalo will also be disrupted in hydrodynamical simulations. %They claim that orphan galaxies should be rarer than expected by a semi-analytical model.

%In this paper, we have discussed the origin and fate of subhalo disruption (orphan) in detail and give a new insight. %Here we give some discussion on the orphan galaxy based on the analysis above. 
Based on hydrodynamical simulations of three cluster halos, \citet{Smith16} proposed a formula to describe the relationship between the bound fraction in the stellar component and the dark matter component,
\begin{equation}
    f_{\text{str}} = 1 - \mathrm{exp} ( - k f_{\text{dm}}),
    \label{eq: fstr}
\end{equation}
where $k$ is a parameter related to the satellite size. With a typical value of $k\sim10$,%\jx{fixme}\fhc{They have three values: 24, 12, 8, depending on the size of the galaxy}
the stellar component is barely stripped even when the dark matter component has been stripped by $90\%$. %\jx{what's the typical $k$? why do you say the following?}This formula suggests that the stellar component embedded within the center of the subhalo is less affected even when the dark matter fraction has been stripped down to $10\%$. In this case, most of the dark matter at the out region of the subhalo will be removed and the satellite inside keeps the status close to its infall status. %\citetalias{Han16} model reveals that the dark matter bound fraction distribution of the subhalo at the $R_{200}$ of the host halo will peak at about $0.4$. The above description is actually a common situation in most satellites in the universe.
However, for very small $f_{\rm dm}$, Equation~\eqref{eq: fstr} predicts $f_{\rm{str}}$ will be proportional to $f_{\rm{dm}}$, so that the stellar mass loss can also be significant when the dark matter becomes heavily stripped. In this work, we have shown that pre-processed subhalos are more likely to lose mass by several orders of magnitude within a short timescale. Consequently, the stellar component of a subhalo will also lose mass by similar orders of magnitude, even though it resides at the center of the subhalo. For example, in Fig.~\ref{fig:satgen_example2}, at the time this subhalo falls below the resolution limit, \satgen model predicts a tidal radius about $0.01h^{-1}\rm{kpc}$, far smaller than its initial halo size and galaxy size. No orphan galaxy will remain at the present time in this subhalo, and the stellar component will merge into the parent halo during this process. Thus, whether an orphan galaxy should be introduced and the size of it will depend on the final bound mass of the subhalo.

The reliability of this inference depends on whether Eq.~\ref{eq: fstr} is still applicable when $f_{\rm{dm}}$ is very small. Hydrodynamical simulations of very high resolution are needed to quantify the formula in the future. It will also be interesting to model the stellar stripping process and the satellite-subhalo connection statistically in the simulations, to guide semi-analytical models on how to model orphan galaxies, as well as to facilitate easy comparisons between simulations and observations. %{\tt\string SatGen} provides a recipe to model the stripping of the stellar component \citetalias{Jiang21}, which is based on the tidal track evolution \citep{Penarrubia05}. However, the recipe for modelling the stellar mass evolution depends on the evolution of the dark matter structure. In principle, it does not predict the emergence of orphan galaxies.

\section{Summary and conclusions}\label{sec:conclusion}
In this work, we investigate the origin of subhalo disruption in cosmological simulations and its impact on subhalo statistics. We make use of the Aquarius simulation, a very high resolution zoom-in simulation of galactic size halos, to categorize subhalos into different evolution classes and investigate the physical properties responsible for their different fates. 

We find that the segregation in the fates of subhalos corresponds to a segregation in their mass loss rates, which can be traced to \emph{two different acquisition channels} of subhalos at a given infall mass, namely early hierarchical accretion and late direct accretion.

%We find that subhalos that become unresolved in our sample have a different formation channel from those remaining resolved at $z=0$ . 
Focusing on a subhalo sample with $10^4-10^5$ particles upon accretion, almost all subhalos accreted after $z=2$ survive to $z=0$. On the other hand, most subhalos accreted at $z=3$ fall below the simulation resolution at $z=0$ to become orphan subhalos (see Fig.~\ref{fig:massratio_Zacc}).% An additional small fraction of subhalos accreted at high redshift fall directly to the center of its host before disruption and are counted as trapped subhalos corresponding to resolved mergers among subhalos.

The different fates of the two major populations are not due to the different evolution time, but because of the different mass loss rates, with the orphan population typically experiencing \emph{extremely large mass loss rates} compared with the surviving ones (see Fig.~\ref{fig:scaled_massloss}).

The different mass loss rates can be attributed to the different physical properties associated with the subhalos and their hosts. At high redshift, the host halo is in its fast growth phase, resulting in a quick decay of the subhalo orbit due to the rapidly growing potential. In addition, subhalos accreted at high redshift are mostly high order subhalos that are accreted into other halos first before merging into the final host. These high order accretions also correspond mostly to major mergers between the subhalo and its direct host, so that significant dynamical friction is expected compared to minor merger systems. Besides, high redshift progenitors tend to have low concentrations. All these effects contribute to boosting the tidal stripping effect on these early accreted subhalos. Using idealized simulations of high redshift major mergers, we verify that an extreme mass loss rate is indeed expected. 

By contrast, subhalos accreted at low redshift are mostly first order subhalos accreted through minor mergers, with a slowly evolving host halo potential, resulting in much lower mass loss rates. The relatively slow mass loss of the resolved population is consistent with previous studies using idealized simulations which claim that CDM subhalos are robust to tidal stripping and will not disrupt~\citep{vdB18a,Errani20,Errani21}. However, the simulation setups in those works do not meet the physical conditions for the orphan subhalos accreted at high redshift in our sample, and thus do not invalidate the extreme mass loss rates for them. 

\citet{vdB18b} also showed that the mass evolution history in simulations will be unreliable if the force and mass resolution is not sufficient, resulting in artificial disruption. For our simulation setup, the force softening is sufficient for the majority of subhalos in our sample, although insufficient mass resolution can still affect the low mass tail of the mass loss curves (see Fig.~\ref{fig:vdB_cirteria}). Despite this, the segregation of subhalo mass loss rates is already clear well above the reliability thresholds, thus robust to numerical effects.

These findings are consistent with previous statistical models on the subhalo population. Through a careful convergence analysis, \citet{Han16} found that about half of CDM subhalos survive with a lognormal distribution in the bound fraction, while the remaining ones are consistent with being disrupted with 0 mass~\citep[see also][]{He23}. Our results suggest that \emph{the orphan population, disrupted or not, are distinct from the surviving ones}, in having very large mass loss rates. These orphan subhalos are thus expected to have only extremely low final masses, consistent with the 0 mass population in the \citet{Han16} model.

Given the expected low final mass, these orphan subhalos barely contaminate the statistical distribution of the final subhalo population. This is because the number of extremely stripped subhalo arising from high mass progenitors is much smaller than the number of mildly stripped low mass subhalos, due to the steep rise of the subhalo mass function towards the low mass end. To assess the influence quantitatively, we further calculate the tidal stripping of each subhalo in our simulation using a semi-analytical model {\tt\string SatGen}. The modeled subhalo mass evolution, free from artificial effects, supports the simulation data. Numerical effects only slightly underestimate the subhalo mass function by no more than $3\%$ for subhalo particle numbers above 300. Only when the number of particles drops to around 100 do artificial effects from discreteness noise dominate, which will significantly compromise the completeness of the surviving subhalo population below this limit (see Fig.~\ref{fig:satgen_massfunction}).

We clarify that the dominance of minor merger at low redshift is a consequence of a fixed infall mass selection in our subhalo sample, because it has been shown that the progenitor mass distribution is universal across redshift~\citep{Dong22}. However, for any mass limited subhalo sample, it is expected that minor mergers dominate low redshift accretion while major mergers dominate at high redshift, due to the lower host halo masses at high redshift, so the classification of the two acquisition modes remains meaningful. A more detailed investigation of the progenitor composition of different orders will be presented in an upcoming work (Jiang et al. 2024).

Our findings are based on the Aquarius zoom-in simulation which has very high mass and force resolutions. For the level-2 run analyzed in this work, it is able to resolve subhalos down to $\sim 10^{-7}$ the host halo mass, beyond the capability of typical cosmological runs. However, the subhalo statistics from the Aquarius simulations have been shown to be in good agreement with other cosmological simulations~\citep{Aquarius,Han16,HBT+}. This enables us to conclude that while numerical effects do affect the poorest resolved subhalos, the bimodal distribution of the bound fraction, as well as the mass distribution of the surviving ones, have been reliably resolved in modern cosmological simulations.

These findings also have implications on the survival of subhalos in other cosmologies. For example, previous works have found that warm dark matter (WDM) subhalos are more vulnerable to tidal disruption, contributing to the lower abundance compared with CDM subhalos~\citep{Benson13, Bose17b, He23}. The enhanced disruption rate has been mainly attributed to the relatively lower concentration of WDM subhalos. However, according to our results, the accretion mode of a subhalo (early hierarchical accretion vs late direct accretion) is also crucial in determining the mass loss rate. It thus remains to be verified which is the main driver for the enhanced tidal stripping of WDM subhalos.

Hydrodynamical cosmological simulations warrant more attention as we seek to understand galaxy evolution in tidal fields. Numerical effects may also impact the reliability of hydrodynamical simulations. On the other hand, the existence of baryons may help to stabilize the subhalo due to the increased central density, or make it more vulnerable in case of strong feedback, with additional dependencies on the detailed sub-grid physics. It is also important to learn whether and when galaxies inside subhalos disrupt. These problems require further investigation in future works.

%% Also note that the akcnowlodgment environment does not support long amounts of text. If you have a lot of people and institutions to acknowledge, do not use this command. Instead, create a new \section{Acknowledgments}.
\begin{acknowledgments}
We thank Fangzhou Jiang for helpful discussions. This work is supported by National Key R\&D Program of China (2023YFA1607800, 2023YFA1607801), 111 project (No.\ B20019), and the science research grants from the China Manned Space Project (No.\ CMS-CSST-2021-A03). We acknowledge the sponsorship from the Yangyang Development Fund. ZZL acknowledges the funding from the European Union’s Horizon 2020 research and innovation programme under the Marie Skłodowska-Curie grant agreement No. 101109759 (“CuspCore”). The computation of this work is done on the \textsc{Gravity} supercomputer at the Department of Astronomy, Shanghai Jiao Tong University. 
\end{acknowledgments}


\vspace{5mm}
%\facilities{HST(STIS), Swift(XRT and UVOT), AAVSO, CTIO:1.3m,


\software{Numpy \citep{Numpy},  
          Matplotlib \citep{Matplotlib}, 
          \textsc{HBT+} \citep{HBT+},
          \textsc{Gadget-4} \citep{gadget4},
          \satgen \citep{Jiang21}
          }



\appendix

\section{Convergence test of the idealized simulation}\label{sec:convergence}
\begin{figure*}
    \centering
    \includegraphics[scale=0.5]{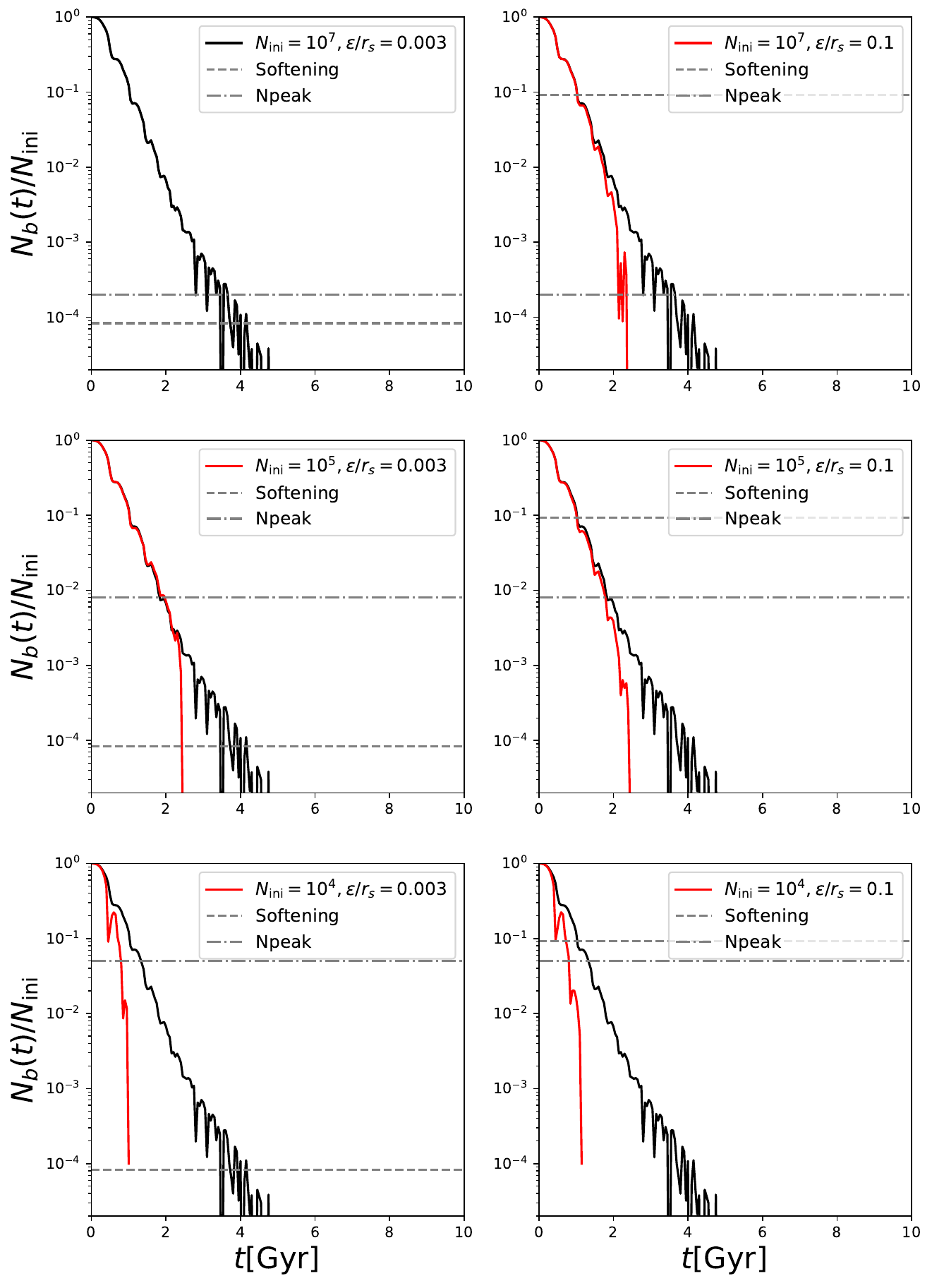}
    \caption{The bound fraction as a function of time for the simulations with different $N_{\rm{ini}}$(different rows) and different softening lengths (different columns), similar to Fig.10 of \citetalias{vdB18b}. All simulations are setup in the same way as in Fig.~\ref{fig:fiducialsnapshot} except for different resolutions. The black line in each panel represents the result of the highest resolution run. The red lines show the results of differing resolutions as labeled. The gray dashed and dash-dotted lines are the bound fraction limits calculated by Eq.~\ref{Eq:soft} and Eq.~\ref{Eq:Npeak}, respectively.}
    \label{fig:soft_Nmax}
\end{figure*}

We follow the procedures of \citetalias{vdB18b} to test the convergence of our idealized simulation in section~\ref{sec:simulationsubhalo}, by rerunning the simulation in different force and mass resolutions. Fig.~\ref{fig:soft_Nmax} shows the bound fraction evolution in each run. In the highest resolution, the subhalo has $N_{\rm{ini}} = 10^{7}$ at the accretion time and the force softening is $\varepsilon/r_s=0.003$, same as the highest resolution used in \citetalias{vdB18b}. 

From the lowest resolution to the highest resolution, we can find that the bound fraction evolution tends to approach that of the highest resolution. Even though we only have a single realization at each resolution, we can gauge the reliable part of the mass resolution curve by checking where it diverges from the highest resolution run. For reference, the horizontal curves show the convergence criteria according to Eq.~\eqref{Eq:soft} and \eqref{Eq:Npeak}. Except for the two lowest resolution runs, the bound fraction can be reliably resolved to a critical value below the more constraining limit of the two. This indicates that the combination of the two limits is conservative for our simulation and subhalo finder combination, and the mass loss curves in our case can be reliably traced down to a slightly lower limit.

\section{Example mass evolution with \satgen}\label{sec:satgen_example}
\begin{figure*}
    \centering
    \begin{subfigure}
        \centering
        \includegraphics[scale=0.6]{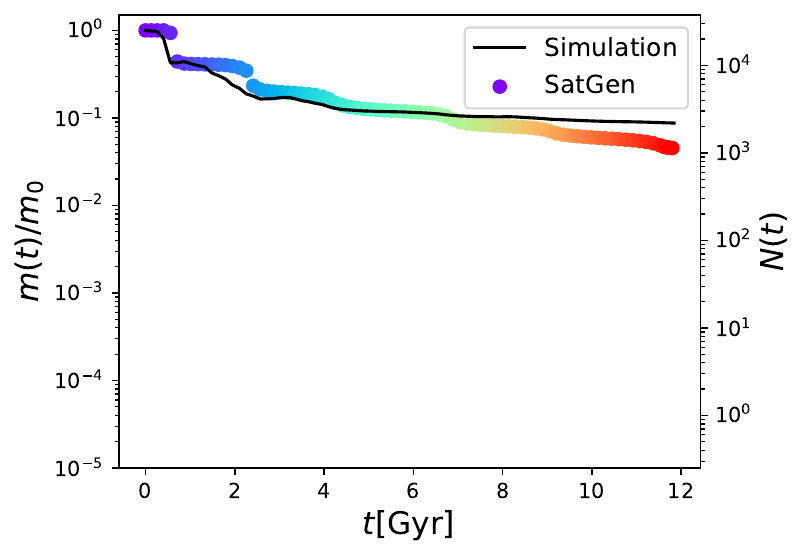}
    \end{subfigure}
    \begin{subfigure}
        \centering
        \includegraphics[scale=0.6]{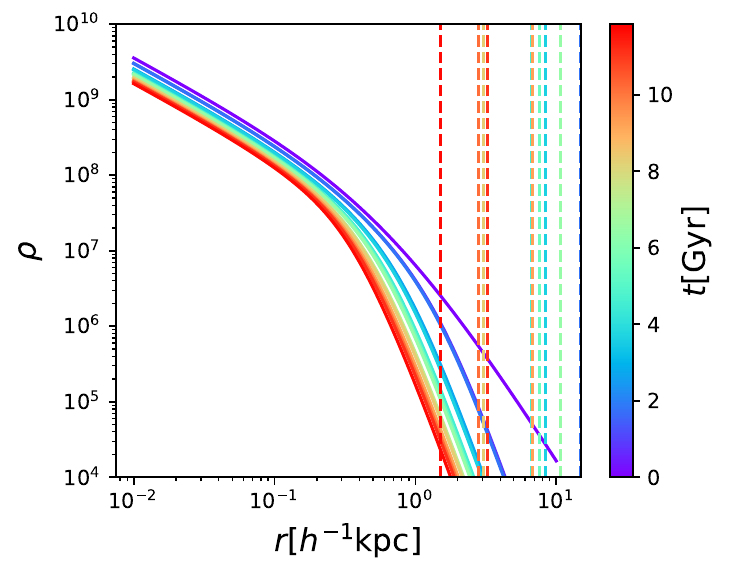}
    \end{subfigure}
    \caption{\textit{Left}: an example mass evolution curve predicted by {\tt\string SatGen} compared with the original simulation. The \satgen data points are color-coded by the evolution time. \textit{Right}: Evolution of the density profile of the same subhalo predicted by \satgen, with the same color-coding as in the left panel. The vertical dashed lines show the instantaneous tidal radius at each snapshot.}
    \label{fig:satgen_example1}
\end{figure*}

We show two examples of how the subhalo evolution module of the {\tt\string SatGen} works on the Aquarius subhalos. Note the mass evolutions are prior to the bias correction we make in Section~\ref{sec:satgen_result}.

The first one is a resolved subhalo case as shown in Fig.~\ref{fig:satgen_example1}. In $N$-body simulation, this subhalo evolves for 12Gyr after it is accreted. It loses about $80\%$ of its mass in the first two orbit periods. After that, it continues to lose mass but at a more modest rate. Finally, this subhalo remains a bound structure with a mass of less than 10\% of its infall mass. The tidal stripping recipe of {\tt\string SatGen} predicts a similar mass evolution. We find that the predicted tidal stripping rate at the pericenter is much stronger than that in the $N$-body simulation. The jumps on the predicted mass loss curve indicate the pericentric passage in the subhalo orbit. Generally, {\tt\string SatGen} removes more mass than the simulation in each period. For most of the evolution time, {\tt\string SatGen} overestimates the stripping efficiency and gives a relatively lower mass. %A corresponding jump happens on the {\tt\string SatGen} mass loss curve at about 8Gyr, which indicates that the mass change still comes from the host tidal field, not the impact of the subhalo intersection.

%Since we do not have the particle data of the Aquarius simulation, 
In the right panel of Fig.~\ref{fig:satgen_example1}, we show the density profile evolution predicted by {\tt\string SatGen}.  The model assumes that the accreted subhalo has an NFW profile. When the subhalo is embedded in the tidal field, its instantaneous density profile is calculated by the instantaneous bound fraction. The corresponding tidal radius shown by the vertical dashed lines evolves slowly in the outer part of the subhalo. At the final time, though the density profile has been sharply truncated, the tidal effect still hardly affects the inner part of the subhalo. 

\begin{figure*}
    \centering
    \begin{subfigure}
        \centering
        \includegraphics[scale=0.6]{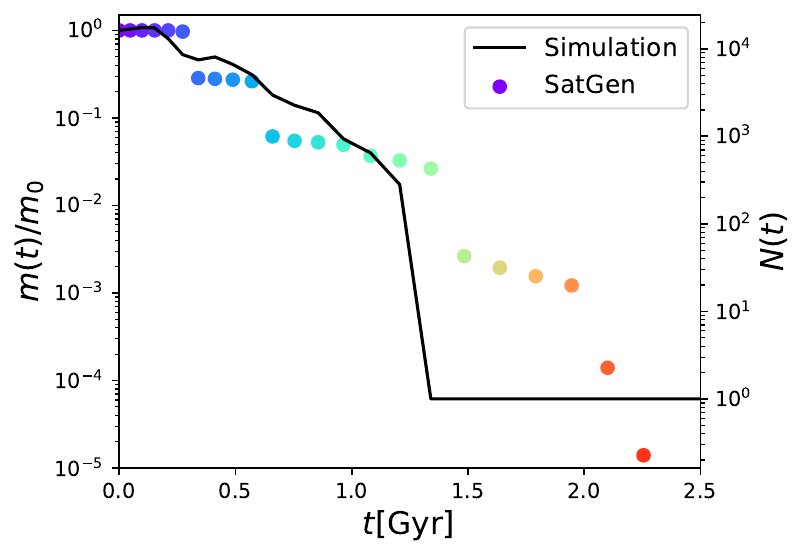}
    \end{subfigure}
    \begin{subfigure}
        \centering
        \includegraphics[scale=0.6]{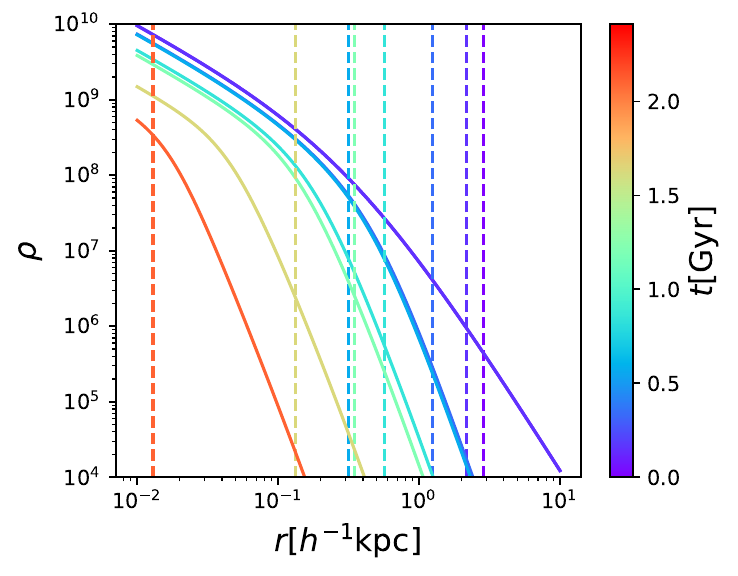}
    \end{subfigure}
    \caption{Same as Fig.~\ref{fig:satgen_example1}, but a showing the result of an orphan subhalo.}
    \label{fig:satgen_example2}
\end{figure*}

The second one is an orphan case as shown in Fig.~\ref{fig:satgen_example2}. In contrast to the resolved case, this subhalo drops below the resolution limit by about 1.2Gyr after accretion. We only have a few tens of snapshots to record its revolution. %We find that this subhalo loses almost all of its material at every pericentric process\jx{this sentence is not making sense}. 
In the $N$-body simulation, it loses all of its mass at the third peri-centric passage. At the snapshot just before its orphan time, it has about 100 particles. Afterwards the subhalo becomes unresolved in the simulation, while \textsc{hbt+} continues to track the motion of its most bound particle. \satgen predicts that this subhalo falls below the resolution limit of $10^{-5}$ set by {\tt\string SatGen} at the fourth peri-centric passage, which is about 2.3Gyr after its accretion time. 

%The density profile evolution gives support. 
As shown in the right panel, this halo has a more extended density profile than the resolved case at the accretion time, while the tidal radius drops quickly over time. At the final time, the tidal radius reduces to about $0.01h^{-1}\rm{kpc}$, and it is very hard to find any bound structure on such a small scale.


\bibliography{sample631}{}
\bibliographystyle{aasjournal}



\end{document}